\documentclass[prd,showkeys,floatfix,twocolumn,amsmath,amssymb,floatfix]{revtex4}
\usepackage{graphicx}
\usepackage{epstopdf}
\usepackage{multirow}
\usepackage{subfigure}

\allowdisplaybreaks[3]

\begin{document}

\title{Identifying the $\Lambda_b(6146)^0$ and $\Lambda_b(6152)^0$ as $D$-wave bottom baryons}

\author{Qiang Mao$^{1}$}
\author{Hua-Xing Chen$^{2,3}$}
\email{hxchen@buaa.edu.cn}
\author{Hui-Min Yang$^2$}

\affiliation{
$^1$Department of Electrical and Electronic Engineering, Suzhou University, Suzhou 234000, China\\
$^2$School of Physics, Beihang University, Beijing 100191, China \\
$^3$School of Physics, Southeast University, Nanjing 210094, China
}

\begin{abstract}
We study the $\Lambda_b(6146)^0$ and $\Lambda_b(6152)^0$ recently observed by LHCb using the method of QCD sum rules within the framework of heavy quark effective theory. Our results suggest that they can be interpreted as $D$-wave bottom baryons of $J^P = 3/2^+$ and $5/2^+$ respectively, both of which contain two $\lambda$-mode excitations. We also investigate other possible assignments containing $\rho$-mode excitations. We extract all the parameters that are necessary to study their decay properties when using the method of light-cone sum rules. We predict masses of their strangeness partners to be $m_{\Xi_b(3/2^+)} = 6.26^{+0.11}_{-0.14}$~GeV and $m_{\Xi_b(5/2^+)} = 6.26^{+0.11}_{-0.14}$~GeV with the mass splitting $\Delta M~=~m_{\Xi_b(5/2^+)} - m_{\Xi_b(3/2^+)} = 4.5^{+1.9}_{-1.5}$~MeV, and propose to search for them in future LHCb and CMS experiments.
\end{abstract}
\maketitle
\pagenumbering{arabic}

\section{Introduction}

In the past few years important experimental progresses were made in the field of bottom baryons. All the $S$-wave singly bottom baryons, except the $\Omega_b^*$ of $J^P = 3/2^+$, have been well observed in experiments~\cite{pdg}. However,  no excited bottom baryons were established until the {LHCb} Collaboration discovered the $\Lambda_b(5912)^0$ and $\Lambda_b(5920)^0$ in 2012~\cite{Aaij:2012da}, which were later confirmed by the {CDF} Collaboration~\cite{Aaltonen:2013tta}. At that time, these were the only two excited bottom baryons well observed in experiments, while in the past two years the LHCb and {CMS} Collaborations continuously observed as many as nine excited bottom baryons:
\begin{itemize}

\item In 2018 the LHCb Collaboration reported their discoveries of two excited bottom baryons, the $\Sigma_{b}(6097)^{\pm}$ in the $\Lambda^0_b \pi^\pm$ invariant mass spectrum and the $\Xi_{b}(6227)^{-}$ in both the $\Lambda^0_b K^-$ and $\Xi^0_b \pi^-$ invariant mass spectra~\cite{Aaij:2018yqz,Aaij:2018tnn};

\item In 2020 the LHCb Collaboration discovered four excited $\Omega_b$ states, $\Omega_b(6316)^-$, $\Omega_b(6330)^-$, $\Omega_b(6340)^-$, and $\Omega_b(6350)^-$, at the same time in the $\Xi_b^0 K^-$ invariant mass spectrum~\cite{Aaij:2020cex};

\item In 2019 the LHCb Collaboration reported their discovery of two excited bottom baryons $\Lambda_b(6146)^0$ and $\Lambda_b(6152)^0$ in the $\Lambda_b^0 \pi^+ \pi^-$ invariant mass distribution~\cite{Aaij:2019amv}:
    \begin{eqnarray}
    \nonumber \Lambda_b(6146)^0 &:& M = 6146.17 \pm 0.33 \pm 0.22 \pm 0.16~{\rm MeV} \, ,
    \\       && \Gamma = 2.9 \pm 1.3 \pm 0.3~{\rm MeV} \, ,
    \\ \nonumber \Lambda_b(6152)^0 &:& M = 6152.51 \pm 0.26 \pm 0.22 \pm 0.16~{\rm MeV} \, ,
    \\       && \Gamma = 2.1 \pm 0.8 \pm 0.3~{\rm MeV} \, .
    \end{eqnarray}
    During this experiment, they observed significant $\Lambda_b(6146)^0 \to \Sigma_b^{*\pm} \pi^\mp$, $\Lambda_b(6152)^0 \to \Sigma_b^\pm \pi^\mp$, and $\Lambda_b(6152)^0 \to \Sigma_b^{*\pm} \pi^\mp$ signals, but no significant $\Lambda_b(6146)^0 \to \Sigma_b^{\pm} \pi^\mp$ signals were observed. The LHCb Collaboration suggested these two states to be the $\Lambda_b(1D)$ baryons, by comparing with the masses predicted by the constituent quark model~\cite{Capstick:1986bm,Ebert:2007nw,Roberts:2007ni,Chen:2014nyo}.

    Later in 2020 the CMS Collaboration confirmed the $\Lambda_b(6146)^0$ and $\Lambda_b(6152)^0$, and measured their masses to be~\cite{Sirunyan:2020gtz}:
    \begin{eqnarray}
    \Lambda_b(6146)^0 &:& M = 6146.5 \pm 1.9 \pm 0.8 \pm 0.2~{\rm MeV} \, ,
    \\ \Lambda_b(6152)^0 &:& M = 6152.7 \pm 1.1 \pm 0.4 \pm 0.2~{\rm MeV} \, .
    \end{eqnarray}
    Besides, they further observed a broad excess of events in the $\Lambda_b^0 \pi^+ \pi^-$ mass distribution in the region of $6040$-$6100$~MeV, whose mass and width were later measured by LHCb to be~\cite{Aaij:2020rkw}:
    \begin{eqnarray}
    \nonumber \Lambda_b(6072)^0 &:& M = 6072.3 \pm 2.9 \pm 0.6 \pm 0.2~{\rm MeV} \, ,
    \\       && \Gamma = 72 \pm 11 \pm 2~{\rm MeV} \, .
    \end{eqnarray}

\end{itemize}

Much earlier, the $\Lambda_b(5912)^0$ and $\Lambda_b(5920)^0$ had been studied by Capstick and Isgur in 1986 as $P$-wave bottom baryons using the relativistic quark model~\cite{Capstick:1986bm}, and their predicted masses are in very good agreement with the LHCb and CDF results obtained in 2012~\cite{Aaij:2012da,Aaltonen:2013tta}. Besides, various phenomenological methods and models were applied to study excited bottom baryons in the past 30 years, such as the constituent quark model~\cite{Garcilazo:2007eh,Ortega:2012cx,Yoshida:2015tia,Gutierrez-Guerrero:2019uwa}, the relativistic quark model~\cite{Ebert:2007nw}, the chiral quark model~\cite{Wang:2018fjm,Kawakami:2019hpp,Xiao:2020oif}, the heavy quark effective theory~\cite{Roberts:2007ni}, the quark pair creation model~\cite{Chen:2018orb,Chen:2018vuc,Yang:2018lzg,Liang:2020hbo}, the relativistic flux tube model~\cite{Chen:2014nyo}, the color hyperfine interaction~\cite{Karliner:2008sv,Karliner:2015ema}, the chiral perturbation theory~\cite{Lu:2014ina,Cheng:2015naa}, and
Lattice {QCD}~\cite{Padmanath:2013bla,Padmanath:2017lng,Burch:2015pka}, etc. These studies are all based on the traditional excited bottom baryon interpretation, while there also exists the molecular interpretation~\cite{GarciaRecio:2012db,Liang:2014eba,An:2017lwg,Montana:2017kjw,Debastiani:2017ewu,Chen:2017xat,Nieves:2017jjx,Huang:2018wgr,Yu:2018yxl,Nieves:2019jhp,Liang:2020dxr}. We refer to recent reviews for detailed discussions~\cite{Chen:2016spr,Cheng:2015iom,Crede:2013sze,Klempt:2009pi,Bianco:2003vb,Korner:1994nh}.

We have systematically investigated mass spectra of excited heavy baryons in~\cite{Chen:2015kpa,Mao:2015gya,Chen:2016phw,Mao:2017wbz} using the method of QCD sum rules~\cite{Shifman:1978bx,Reinders:1984sr} within the framework of Heavy Quark Effective Theory (HQET)~\cite{Grinstein:1990mj,Eichten:1989zv,Falk:1990yz}. More studies on heavy mesons and baryons containing a single heavy quark can be found in~\cite{Bagan:1991sg,Neubert:1991sp,Broadhurst:1991fc,Ball:1993xv,Huang:1994zj,Dai:1996yw,Colangelo:1998ga,Groote:1996em,Zhu:2000py,Lee:2000tb,Huang:2000tn,Wang:2003zp,Duraes:2007te,Liu:2007fg,Zhou:2014ytp,Zhou:2015ywa,Aliev:2018vye,Aliev:2018lcs,Wang:2020pri}.
Our results suggest that the eight excited bottom baryons, $\Lambda_b(5912)^0$, $\Lambda_b(5920)^0$, $\Sigma_{b}(6097)^{\pm}$, $\Xi_{b}(6227)^{-}$, $\Omega_b(6316)^-$, $\Omega_b(6330)^-$, $\Omega_b(6340)^-$, and $\Omega_b(6350)^-$, can be well explained as $P$-wave bottom baryons~\cite{Mao:2015gya,Cui:2019dzj,Chen:2020mpy}.

In this paper we shall use the same approach to study $D$-wave bottom baryons. Some of these studies have been done in our previous papers~\cite{Chen:2016phw,Mao:2017wbz}, but at that time: (a) We did not construct all the bottom baryon interpolating fields, and (b) we did not complete all the sum rule calculations. In the present study we shall finish these two steps and systematically study $D$-wave bottom baryons of the $SU(3)$ flavor $\mathbf{\bar 3}_F$. The obtained results will be used to examine whether the $\Lambda_b(6146)^0$ and $\Lambda_b(6152)^0$ can be interpreted as $D$-wave bottom baryons. Before doing this, we note that this assignment has been discussed and supported by several theoretical studies, using the chiral quark model~\cite{Wang:2019uaj}, the quark pair creation model~\cite{Liang:2019aag,Chen:2019ywy}, and QCD sum rules~\cite{Azizi:2020tgh}, etc.

This paper is organized as follows. In Section~\ref{sec:current}, we construct all the interpolating fields for $D$-wave bottom baryons of the $SU(3)$ flavor $\mathbf{\bar 3}_F$, which are used to perform QCD sum rule analyses in Section~\ref{sec:sumrule}. The obtained sum rule equations are further used to perform numerical analyses in Section~\ref{sec:numerical}. In Section~\ref{sec:summary} we discuss the results and conclude this paper.

\section{Interpolating fields for the $D$-wave bottom baryon}
\label{sec:current}

\begin{figure*}[htb]
\begin{center}
\scalebox{0.6}{\includegraphics{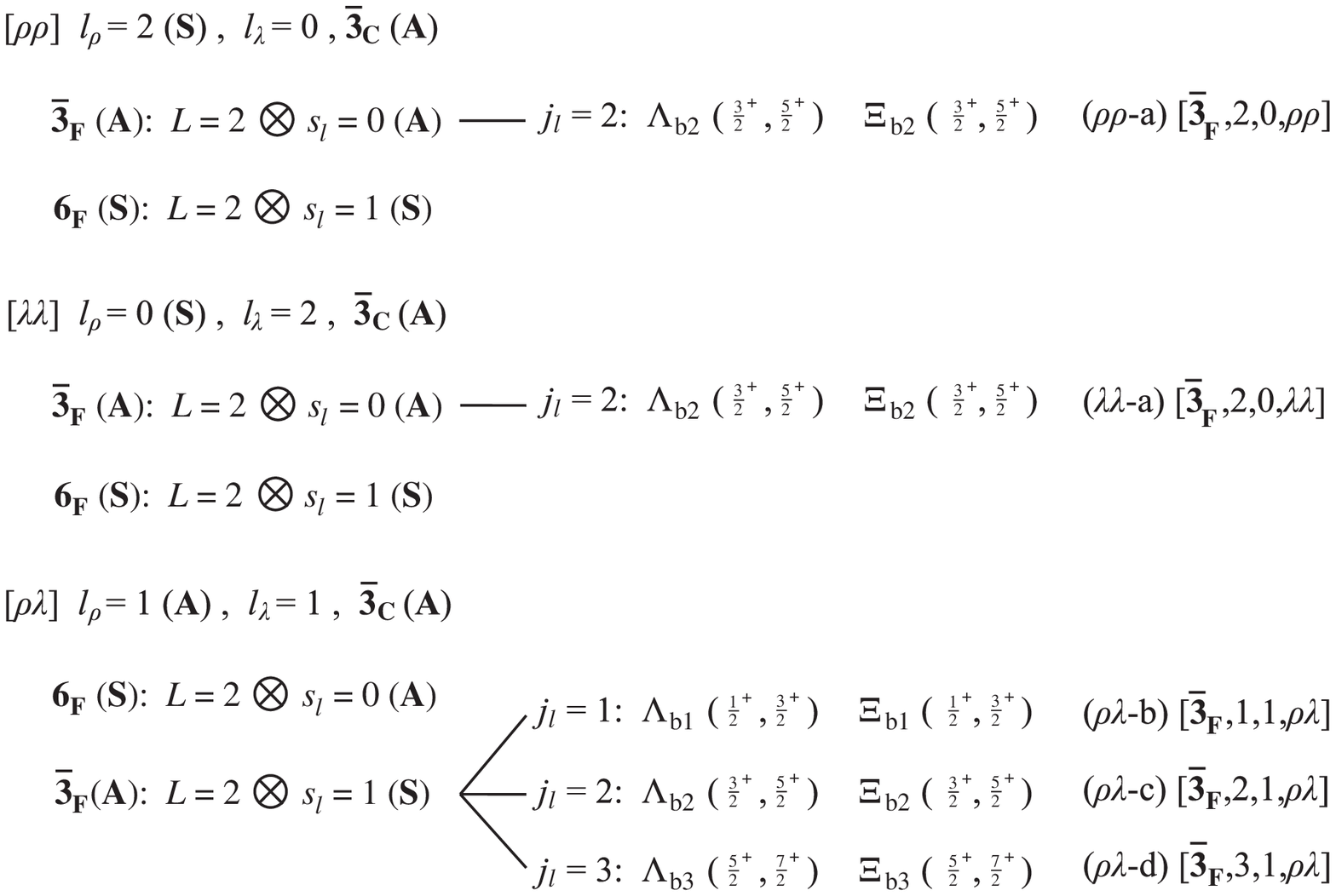}}
\end{center}
\caption{Categorization of $D$-wave bottom baryons belonging to the $SU(3)$ flavor $\mathbf{\bar3}_F$ representation.
\label{fig:dwave}}
\end{figure*}

The $D$-waves heavy baryons have been systematically classified in~\cite{Chen:2007xf}, and their interpolating fields have been partly constructed in~\cite{Chen:2016phw,Mao:2017wbz}. In this section we further construct all the $D$-wave heavy baryon interpolating fields of the $SU(3)$ flavor $\mathbf{\bar3}_F$. Note that some of them are different from those given in~\cite{Chen:2016phw,Mao:2017wbz}, since we have explicitly used several projection operators in the present~study.

First we briefly introduce our notations. A $D$-wave bottom baryon consists of one $bottom$ quark and two light $up/down/strange$ quarks. We use $l_\rho$ to denote the orbital angular momentum between the two light quarks, and $l_\lambda$ to denote the orbital angular momentum between the bottom quark and the two-light-quark system. There can be $\rho\rho$-mode excited $D$-wave bottom baryons ($l_\rho = 2$ and $l_\lambda = 0$ into $L = 2$), $\lambda\lambda$-mode ones ($l_\rho = 0$ and $l_\lambda = 2$ into $L = 2$), and $\rho\lambda$-mode ones ($l_\rho = 1$ and $l_\lambda = 1$ into $L = 2$). Altogether its internal symmetries are as follows:
\begin{itemize}

\item Color structure of the two light quarks is antisymmetric ($\mathbf{\bar 3}_C$);

\item Flavor structure of the two light quarks is either antisymmetric ($\mathbf{\bar 3}_F$) or symmetric ($\mathbf{6}_F$);

\item Spin structure of the two light quarks is either antisymmetric ($s_l = 0$) or symmetric ($s_l = 1$);

\item Orbital structure of the two light quarks is either antisymmetric ($l_\rho = 1$) or symmetric ($l_\rho = 0/2$);

\item Totally, the two light quarks are antisymmetric due to the Pauli principle.

\end{itemize}
Accordingly, we categorize $D$-wave bottom baryons into 12 multiplets, five of which belong to the $SU(3)$ flavor $\mathbf{\bar 3}_F$ representation, as shown in Figure~\ref{fig:dwave}. We denote them as $[F(lavor), j_l, s_l, \rho/\lambda]$, where $j_l$ is the total angular momentum of the light components ($j_l = l_\lambda \otimes l_\rho \otimes s_l$). Each multiplet contains two bottom baryons, whose total angular momentum are $j = j_l \otimes s_b = j_l \pm 1/2$, with $s_b$ the spin of the bottom quark.

We use the notation $J^{\alpha_1\cdots\alpha_{j-1/2}}_{j,P,F,j_l,s_l,\rho/\lambda}$ to denote the $D$-wave bottom baryon interpolating field, and~separately construct them for the $[\mathbf{\bar 3}_F, 2, 0, \rho\rho]$, $[\mathbf{\bar 3}_F, 2, 0, \lambda\lambda]$, $[\mathbf{\bar 3}_F, 1, 1, \rho\lambda]$, $[\mathbf{\bar 3}_F, 2, 1, \rho\lambda]$, and $[\mathbf{\bar 3}_F, 3, 1, \rho\lambda]$ multiplets. Note that Equations~(\ref{eq:current1}), (\ref{eq:current2}), (\ref{eq:current3}), (\ref{eq:current4}), and (\ref{eq:current10}) are the same as those given in~\cite{Chen:2016phw,Mao:2017wbz} except for some overall factors; Equations~(\ref{eq:current5}), (\ref{eq:current6}), and (\ref{eq:current9}) are different since we have explicitly used some projection operators in the present study; and  Equations~(\ref{eq:current7}) and (\ref{eq:current8}) were not constructed in~\cite{Chen:2016phw,Mao:2017wbz}. Note that we need to use certain projection operators to distinguish the rich internal structures of $D$-wave bottom baryons. In~\cite{Chen:2016phw,Mao:2017wbz} we constructed some of them, but there we could only project baryons into definite total angular momenta (spins). In the present study we  constructed all  operators to project baryons into definite internal angular momenta~(spins).
\begin{widetext}
\begin{itemize}

\item The bottom baryon doublet $[\mathbf{\bar 3}_F, 2, 0, \rho\rho]$ contains two bottom baryons of $j^P = 3/2^+$ and $5/2^+$, whose interpolating fields are:
\begin{eqnarray}
&& J^{\alpha}_{3/2,+,\mathbf{\bar 3}_F,2,0,\rho\rho}(x)
\label{eq:current1}
\\ \nonumber &=& \epsilon_{abc} \Big ( [\mathcal{D}^t_{\mu_1} \mathcal{D}^t_{\mu_2} q^{aT}(x)] \mathbb{C} \gamma_5 q^b(x) - 2 [\mathcal{D}^t_{\mu_1} q^{aT}(x)] \mathbb{C} \gamma_5 [\mathcal{D}^t_{\mu_2} q^b(x)] + q^{aT}(x) \mathbb{C} \gamma_5 [\mathcal{D}^t_{\mu_1} \mathcal{D}^t_{\mu_2} q^b(x)] \Big )
\\ \nonumber && ~~~~~~~~~~~~~~~~~~~~~~~~~~~~~~~~~~~~~~~~~~~~~~~~~~~~~~~~~~~~~~~~~~~~~~~~~~~~~~~~~
\times \Gamma_{J=2}^{\alpha\mu_4,\mu_1\mu_2} \times \gamma^t_{\mu_4} \gamma_5 h_v^c(x) \, ,
\\ && J^{\alpha_1\alpha_2}_{5/2,+,\mathbf{\bar 3}_F,2,0,\rho\rho}(x)
\label{eq:current2}
\\ \nonumber &=& \epsilon_{abc} \Big ( [\mathcal{D}^t_{\mu_1} \mathcal{D}^t_{\mu_2} q^{aT}(x)] \mathbb{C} \gamma_5 q^b(x) - 2 [\mathcal{D}^t_{\mu_1} q^{aT}(x)] \mathbb{C} \gamma_5 [\mathcal{D}^t_{\mu_2} q^b(x)] + q^{aT}(x) \mathbb{C} \gamma_5 [\mathcal{D}^t_{\mu_1} \mathcal{D}^t_{\mu_2} q^b(x)] \Big )
\\ \nonumber && ~~~~~~~~~~~~~~~~~~~~~~~~~~~~~~~~~~~~~~~~~~~~~~~~~~~~~~~~~~~~~~~~~~~~~~~~~~~~~~~~~~~~~~~
\times \Gamma_{J=5/2}^{\alpha_1\alpha_2,\mu_1\mu_2} \times  h_v^c(x) \, .
\end{eqnarray}
In the above expressions: $a,b,c$ are color indices, and $\epsilon_{abc}$ is the totally antisymmetric tensor; $\mathbb{C}$ is the charge-conjugation operator; $q(x)$ is the light $up/down/strange$ quark field, and $h_v(x)$ is the $bottom$ quark field; $\gamma^t_\mu = \gamma_\mu - v\!\!\!\slash v_\mu$, $\mathcal{D}_\mu = \partial_\mu - i g A_\mu$, $\mathcal{D}_\mu^t = \mathcal{D}_\mu - (\mathcal{D} \cdot v) v_\mu$, $g_t^{\alpha_1\alpha_2}=g^{\alpha_1\alpha_2} - v^{\alpha_1} v^{\alpha_2}$, and $v$ is the velocity of the bottom quark; and  $\Gamma_{J=2}^{\alpha\beta,\mu\nu}$ and $\Gamma_{J=5/2}^{\alpha\beta,\mu\nu}$ are the $J = 2$ and $J = 5/2$ projection~operators:
\begin{eqnarray}
\Gamma_{J=2}^{\alpha\beta,\mu\nu} &=& g_t^{\alpha \mu} g_t^{\beta \nu} + g_t^{\alpha \nu} g_t^{\beta \mu} - {2\over3}~g_t^{\alpha \beta} g_t^{\mu \nu} \, ,
\\
\Gamma_{J=5/2}^{\alpha\beta,\mu\nu} &=& g_t^{\alpha\mu} g_t^{\beta\nu} + g_t^{\alpha\nu} g_t^{\beta\mu} - {2 \over 5}~g_t^{\alpha\beta} g_t^{\mu\nu}
 - {1 \over 5}~g_t^{\alpha\mu} \gamma_t^{\beta}\gamma_t^{\nu} - {1 \over 5}~g_t^{\alpha\nu} \gamma_t^{\beta}\gamma_t^{\mu}
 - {1 \over 5}~g_t^{\beta\mu} \gamma_t^{\alpha}\gamma_t^{\nu} - {1 \over 5}~g_t^{\beta\nu} \gamma_t^{\alpha}\gamma_t^{\mu} \, .
\end{eqnarray}

\item The bottom baryon doublet $[\mathbf{\bar 3}_F, 2, 0, \lambda\lambda]$ contains two bottom baryons of $j^P = 3/2^+$ and $5/2^+$, whose interpolating fields are:
\begin{eqnarray}
&& J^{\alpha}_{3/2,+,\mathbf{\bar 3}_F,2,0,\lambda\lambda}(x)
\label{eq:current3}
\\ \nonumber &=& \epsilon_{abc} \Big ( [\mathcal{D}^t_{\mu_1} \mathcal{D}^t_{\mu_2} q^{aT}(x)] \mathbb{C} \gamma_5 q^b(x) + 2 [\mathcal{D}^t_{\mu_1} q^{aT}(x)] \mathbb{C} \gamma_5 [\mathcal{D}^t_{\mu_2} q^b(x)] + q^{aT}(x) \mathbb{C} \gamma_5 [\mathcal{D}^t_{\mu_1} \mathcal{D}^t_{\mu_2} q^b(x)] \Big )
\\ \nonumber && ~~~~~~~~~~~~~~~~~~~~~~~~~~~~~~~~~~~~~~~~~~~~~~~~~~~~~~~~~~~~~~~~~~~~~~~~~~~~~~~~~
\times \Gamma_{J=2}^{\alpha\mu_4,\mu_1\mu_2} \times \gamma^t_{\mu_4} \gamma_5 h_v^c(x) \, ,
\\ && J^{\alpha_1\alpha_2}_{5/2,+,\mathbf{\bar 3}_F,2,0,\lambda\lambda}(x)
\label{eq:current4}
\\ \nonumber &=& \epsilon_{abc} \Big ( [\mathcal{D}^t_{\mu_1} \mathcal{D}^t_{\mu_2} q^{aT}(x)] \mathbb{C} \gamma_5 q^b(x) + 2 [\mathcal{D}^t_{\mu_1} q^{aT}(x)] \mathbb{C} \gamma_5 [\mathcal{D}^t_{\mu_2} q^b(x)] + q^{aT}(x) \mathbb{C} \gamma_5 [\mathcal{D}^t_{\mu_1} \mathcal{D}^t_{\mu_2} q^b(x)] \Big )
\\ \nonumber && ~~~~~~~~~~~~~~~~~~~~~~~~~~~~~~~~~~~~~~~~~~~~~~~~~~~~~~~~~~~~~~~~~~~~~~~~~~~~~~~~~~~~~~~
\times \Gamma_{J=5/2}^{\alpha_1\alpha_2,\mu_1\mu_2} \times  h_v^c(x) \, .
\end{eqnarray}

\item The bottom baryon doublet $[\mathbf{\bar 3}_F, 1, 1, \rho\lambda]$ contains two bottom baryons of $j^P = 1/2^+$ and $3/2^+$, whose interpolating fields are:
\begin{eqnarray}
J_{1/2,+,\mathbf{\bar 3}_F,1,1,\rho\lambda}(x)
&=& \epsilon_{abc} \Big ( [\mathcal{D}^t_{\mu_1} \mathcal{D}^t_{\mu_2} q^{aT}(x)] \mathbb{C} \gamma^t_{\mu_3} q^b(x) - q^{aT}(x) \mathbb{C} \gamma^t_{\mu_3} [\mathcal{D}^t_{\mu_1} \mathcal{D}^t_{\mu_2} q^b(x)] \Big )
\label{eq:current5}
\\ \nonumber && ~~~~~~~~~~~~~~~~~~~~~~~~~~~~~~~~~~~~~~~~~~~~~~~~~~~~~~~~~\,
\times \Gamma_{J=2}^{\mu_3\mu_4,\mu_1\mu_2} \times \gamma^t_{\mu_4} \gamma_5 h_v^c(x) \, ,
\\ J^{\alpha}_{3/2,+,\mathbf{\bar 3}_F,1,1,\rho\lambda}(x)
&=& \epsilon_{abc} \Big ( [\mathcal{D}^t_{\mu_1} \mathcal{D}^t_{\mu_2} q^{aT}(x)] \mathbb{C} \gamma^t_{\mu_3} q^b(x) - q^{aT}(x) \mathbb{C} \gamma^t_{\mu_3} [\mathcal{D}^t_{\mu_1} \mathcal{D}^t_{\mu_2} q^b(x)] \Big )
\label{eq:current6}
\\ \nonumber && ~~~~~~~~~~~~~~~~~~~~~~~~~~~~~~~~~~~~~~~~~~~~~~~~~
\times {\Gamma_{J=3/2,\mu_4}^{\alpha}} \times \Gamma_{J=2}^{\mu_3\mu_4,\mu_1\mu_2} \times h_v^c(x) \, ,
\end{eqnarray}
where $\Gamma_{J=3/2}^{\alpha,\mu}$ is the $J = 3/2$ projection operator:
\begin{eqnarray}
\Gamma_{J=3/2}^{\alpha,\mu} &=& g_t^{\alpha\mu} - {1 \over 3}~\gamma_t^{\alpha} \gamma_t^{\mu} \, .
\end{eqnarray}

\item The bottom baryon doublet $[\mathbf{\bar 3}_F, 2, 1, \rho\lambda]$ contains two bottom baryons of $j^P = 3/2^+$ and $5/2^+$, whose interpolating fields are:
\begin{eqnarray}
J^{\alpha}_{3/2,+,\mathbf{\bar 3}_F,2,1,\rho\lambda}(x)
\label{eq:current7}
&=& \epsilon_{abc} \Big ( [\mathcal{D}^t_{\mu_1} \mathcal{D}^t_{\mu_2} q^{aT}(x)] \mathbb{C} \gamma^t_{\mu_3} q^b(x) - q^{aT}(x) \mathbb{C} \gamma^t_{\mu_3} [\mathcal{D}^t_{\mu_1} \mathcal{D}^t_{\mu_2} q^b(x)] \Big )
\\ \nonumber && ~~~~~~~~~~~~~~~~~~~~~~~~~~~~~~~~~~
\times {\Gamma_{J=3/2,\mu_5}^{\alpha}} \times \Gamma_{J=2}^{\mu_4\mu_5,\mu_1\mu_2} \times (-2i)~\sigma^{\mu_3\mu_4} h_v^c(x) \, ,
\\ J^{\alpha_1\alpha_2}_{5/2,+,\mathbf{\bar 3}_F,2,1,\rho\lambda}(x)
\label{eq:current8}
&=& \epsilon_{abc} \Big ( [\mathcal{D}^t_{\mu_1} \mathcal{D}^t_{\mu_2} q^{aT}(x)] \mathbb{C} \gamma^t_{\mu_3} q^b(x) - q^{aT}(x) \mathbb{C} \gamma^t_{\mu_3} [\mathcal{D}^t_{\mu_1} \mathcal{D}^t_{\mu_2} q^b(x)] \Big )
\\ \nonumber && ~~~~~~~~~~~~~~~~~~~~~~~~
\times {\Gamma_{J=5/2,\mu_5\mu_6}^{\alpha_1\alpha_2}} \times \Gamma_{J=2}^{\mu_4\mu_5,\mu_1\mu_2} \times \epsilon^{\mu_3\mu_4\mu_6\mu_9}  \times \gamma^t_{\mu_9} \gamma_5 h_v^c(x) \, .
\end{eqnarray}

\item The bottom baryon doublet $[\mathbf{\bar 3}_F, 3, 1, \rho\lambda]$ contains two bottom baryons of $j^P = 5/2^+$ and $7/2^+$, whose interpolating fields are:
\begin{eqnarray}
J^{\alpha_1\alpha_2}_{5/2,+,\mathbf{\bar 3}_F,3,1,\rho\lambda}(x)
\label{eq:current9}
&=& \epsilon_{abc} \Big ( [\mathcal{D}^t_{\mu_1} \mathcal{D}^t_{\mu_2} q^{aT}(x)] \mathbb{C} \gamma^t_{\mu_3} q^b(x) - q^{aT}(x) \mathbb{C} \gamma^t_{\mu_3} [\mathcal{D}^t_{\mu_1} \mathcal{D}^t_{\mu_2} q^b(x)] \Big )
\\ \nonumber && ~~~~~~~~~~~~~~~~~~~~~~~~~~~~~~~~~~~~~~~~~~~~~~~~~~~~\,
\times \Gamma_{J=3}^{\alpha_1\alpha_2\mu_4,\mu_1\mu_2\mu_3} \times \gamma^t_{\mu_4} \gamma_5 h_v^c(x) \, ,
\\ J^{\alpha_1\alpha_2\alpha_3}_{7/2,+,\mathbf{\bar 3}_F,3,1,\rho\lambda}(x)
\label{eq:current10}
&=& \epsilon_{abc} \Big ( [\mathcal{D}^t_{\mu_1} \mathcal{D}^t_{\mu_2} q^{aT}(x)] \mathbb{C} \gamma^t_{\mu_3} q^b(x) - q^{aT}(x) \mathbb{C} \gamma^t_{\mu_3} [\mathcal{D}^t_{\mu_1} \mathcal{D}^t_{\mu_2} q^b(x)] \Big )
\\ \nonumber && ~~~~~~~~~~~~~~~~~~~~~~~~~~~~~~~~~~~~~~~~~~~~~~~~~~~~~~~~~~~\,
\times \Gamma_{J=7/2}^{\alpha_1\alpha_2\alpha_3,\mu_1\mu_2\mu_3} \times h_v^c(x) \, ,
\end{eqnarray}
where $\Gamma_{J=3}^{\alpha_1\alpha_2\alpha_3,\mu_1\mu_2\mu_3}$ and $\Gamma_{J=7/2}^{\alpha_1\alpha_2\alpha_3,\mu_1\mu_2\mu_3}$ are the $J = 3$ and $J = 7/2$ projection operators, with $\mathbb{S}^{\prime\prime} [\cdots]$ the symmetrization (only symmetrization but no subtraction) of the trace terms in the sets $(\mu_1 \mu_2 \mu_3)$ and $(\nu_1\nu_2\nu_3)$:
\begin{eqnarray}
\Gamma_{J=3}^{\mu_1\mu_2\mu_3,\nu_1\nu_2\nu_3} &=& \mathbb{S}^{\prime\prime} \Big[ g_t^{\mu_1\nu_1} g_t^{\mu_2\nu_2} g_t^{\mu_3\nu_3}
- {2\over5} ~ g_t^{\mu_1\nu_1} g_t^{\mu_2\mu_3} g_t^{\nu_2\nu_3} \Big] \, ,
\\ \Gamma_{J=7/2}^{\mu_1\mu_2\mu_3,\nu_1\nu_2\nu_3} &=& \mathbb{S}^{\prime\prime} \Big[ g_t^{\mu_1\nu_1} g_t^{\mu_2\nu_2} g_t^{\mu_3\nu_3}
- {3\over7} ~ g_t^{\mu_1\nu_1} g_t^{\mu_2\mu_3} g_t^{\nu_2\nu_3}
+ {2\over35} ~ g_t^{\mu_1\mu_2} g_t^{\nu_1\nu_2} \gamma_t^{\mu_3}\gamma_t^{\nu_3}
- {1\over28} ~ g_t^{\mu_1\nu_1} \gamma_t^{\mu_2} \gamma_t^{\nu_2} \gamma_t^{\mu_3} \gamma_t^{\nu_3} \Big] \, .
\end{eqnarray}

\end{itemize}
\end{widetext}

\section{QCD Sum Rule Analyses}
\label{sec:sumrule}

In this section we use the $D$-wave bottom baryon interpolating field $J^{\alpha_1\cdots\alpha_{j-1/2}}_{j,P,F,j_l,s_l,\rho/\lambda}$ to perform QCD sum rule analyses within the framework of heavy quark effective theory. Because identical sum rules are obtained using both $J^{\alpha_1\cdots\alpha_{j_l-1}}_{j_l-1/2,P,F,j_l,s_l,\rho/\lambda}$ and $J^{\alpha_1\cdots\alpha_{j_l}}_{j_l+1/2,P,F,j_l,s_l,\rho/\lambda}$ within the same multiplet, we only need to use one of them to perform QCD sum rule analyses. In the present study we study the $[\mathbf{\bar 3}_F,2,0,\rho\rho]$, $[\mathbf{\bar 3}_F,2,0,\lambda\lambda]$, $[\mathbf{\bar 3}_F,1,1,\rho\lambda]$, $[\mathbf{\bar 3}_F,2,1,\rho\lambda]$, and $[\mathbf{\bar 3}_F,3,1,\rho\lambda]$ multiplets through the interpolating fields $J^{\alpha}_{3/2,+,\mathbf{\bar 3}_F,2,0,\rho\rho}$, $J^{\alpha}_{3/2,+,\mathbf{\bar 3}_F,2,0,\lambda\lambda}$, $J_{1/2,+,\mathbf{\bar 3}_F,1,1,\rho\lambda}$, $J^{\alpha}_{3/2,+,\mathbf{\bar 3}_F,2,1,\rho\lambda}$, and $J^{\alpha_1\alpha_2}_{5/2,+,\mathbf{\bar 3}_F,3,1,\rho\lambda}$, respectively.

We assume that the interpolating field $J^{\alpha_1\cdots\alpha_{j-1/2}}_{j,P,F,j_l,s_l,\rho/\lambda}$ couples to the bottom baryon belonging to the $[F, j_l, s_l, \rho/\lambda]$ multiplet through:
\begin{eqnarray}
\nonumber && \langle 0| J^{\alpha_1\cdots\alpha_{j-1/2}}_{j,P,F,j_l,s_l,\rho\rho/\lambda\lambda/\rho\lambda} |j,P,F,j_l,s_l,\rho\rho/\lambda\lambda/\rho\lambda \rangle
\\  &=& f_{F,j_l,s_l,\rho\rho/\lambda\lambda/\rho\lambda} u^{\alpha_1\cdots\alpha_{j-1/2}} \, .
\label{eq:coupling}
\end{eqnarray}
Then we can extract the baryon mass to be:
\begin{equation}
\label{eq:mass}
m_{j,P,F,j_l,s_l,\rho/\lambda} = m_b + \overline{\Lambda}_{F,j_l,s_l,\rho/\lambda} + \delta m_{j,P,F,j_l,s_l,\rho/\lambda} \, ,
\end{equation}
where $\overline{\Lambda}_{F,j_l,s_l,\rho/\lambda} = \overline{\Lambda}_{j_l-1/2,P,F,j_l,s_l,\rho/\lambda} = \overline{\Lambda}_{j_l+1/2,P,F,j_l,s_l,\rho/\lambda}$ is the sum rule result at the leading order, and $\delta m_{j,P,F,j_l,s_l,\rho/\lambda}$ is the sum rule result at the ${\mathcal O}(1/m_b)$ order:
\begin{eqnarray}
\label{eq:masscorrection}
&& \delta m_{j,P,F,j_l,s_l,\rho/\lambda} 
\\ \nonumber &=& -\frac{1}{4m_{b}}(K_{F,j_l,s_l,\rho/\lambda} + d_{j,j_{l}}C_{mag}\Sigma_{F,j_l,s_l,\rho/\lambda} ) \, .
\end{eqnarray}
Here $C_{mag} = [ \alpha_s(m_b) / \alpha_s(\mu) ]^{3/\beta_0}$ with $\beta_0 = 11 - 2 n_f /3$, and the coefficient $d_{j,j_{l}}$ is
\begin{equation}
d_{j_{l}-1/2,j_{l}} = 2j_{l}+2\, ,~~~~~
d_{j_{l}+1/2,j_{l}} = -2j_{l} \, .
\end{equation}
Hence, the $\Sigma_{F,j_l,s_l,\rho/\lambda}$ term is directly related to the mass splitting within the same multiplet:
\begin{eqnarray}
&& \Delta M_{F,j_l,s_l,\rho\rho/\lambda\lambda/\rho\lambda} 
\\ \nonumber &=& m_{j_l+1/2,P,F,j_l,s_l,\rho/\lambda} - m_{j_l-1/2,P,F,j_l,s_l,\rho/\lambda} \, .
\end{eqnarray}

As an example, we use the bottom baryon doublet $[\Xi_b(\mathbf{\bar 3}_F),3,1,\rho\lambda]$ to perform QCD sum rule analyses, through the field:
\begin{eqnarray}
&& J^{\alpha_1\alpha_2}_{5/2,+,\Xi_b,3,1,\rho\lambda}(x)
\label{eq:current}
\\ \nonumber &=& \epsilon_{abc} \Big ( [\mathcal{D}^t_{\mu_1} \mathcal{D}^t_{\mu_2} u^{aT}(x)] \mathbb{C} \gamma^t_{\mu_3} s^b(x)
\\ \nonumber && ~~~~~~~~~~~~~~~~~~~~~~~~~~ - u^{aT}(x) \mathbb{C} \gamma^t_{\mu_3} [\mathcal{D}^t_{\mu_1} \mathcal{D}^t_{\mu_2} s^b(x)] \Big )
\\ \nonumber && ~~~~~~~~~~~~~~~~~~~~~~~ \times \Gamma_{J=3}^{\alpha_1\alpha_2\mu_4,\mu_1\mu_2\mu_3} \times \gamma^t_{\mu_4} \gamma_5 h_v^c(x) \, .
\end{eqnarray}
From this field, we obtain:
\begin{eqnarray}
&& \Pi_{\Xi_b,3,1,\rho\lambda} (\omega_c, T) = f_{\Xi_b,3,1,\rho\lambda}^{2} e^{-2 \bar \Lambda_{\Xi_b,3,1,\rho\lambda} / T}
\label{eq:ope}
\\ \nonumber &=& \int_{2m_s}^{\omega_c} [ \frac{1}{12800\pi^4} \omega^9 - \frac{3m_s^2}{800\pi^4} \omega^7 + \frac{63m_s\langle \bar s s \rangle}{800\pi^2} \omega^5
\\ \nonumber && - \frac{21m_s\langle \bar q q \rangle}{400\pi^2} \omega^5 - \frac{21\langle g_s^2 GG \rangle}{3200\pi^4} \omega^5 + \frac{21m_s\langle g_s \bar q \sigma Gq \rangle}{80\pi^2} \omega^3
\\ \nonumber && + \frac{63m_s^2\langle g_s^2 GG \rangle}{1280\pi^4}\omega^3 - \frac{63m_s\langle g_s^2 GG \rangle\langle \bar s s \rangle}{320\pi^2}\omega ]e^{-\omega/T}d\omega \, ,
\\ && f_{\Xi_b,3,1,\rho\lambda}^{2} K_{\Xi_b,3,1,\rho\lambda} e^{-2 \bar \Lambda_{\Xi_b,3,1,\rho\lambda} / T}
\label{eq:Kc}
\\ \nonumber &=& \int_{2m_s}^{\omega_c} [ -\frac{223}{9856000\pi^4} \omega^{11} + \frac{1759m_s^2}{1209600\pi^4} \omega^9
\\ \nonumber &&  - \frac{543m_s\langle \bar s s \rangle}{11200\pi^2} \omega^7 + \frac{17m_s\langle \bar q q \rangle}{672\pi^2} \omega^7 +\frac{80147\langle g_s^2 GG \rangle}{38707200\pi^4} \omega^7
\\ \nonumber &&  - \frac{181m_s\langle g_s \bar q \sigma Gq \rangle}{800\pi^2} \omega^5 - \frac{1091m_s^2\langle g_s^2 GG \rangle}{38400\pi^4} \omega^5
\\ \nonumber &&  - \frac{307m_s\langle g_s^2 GG \rangle\langle \bar q q \rangle}{4320\pi^2}\omega^3+ \frac{779m_s\langle g_s^2 GG \rangle\langle \bar s s \rangle}{3456\pi^2} \omega^3
\\ \nonumber &&  + \frac{59m_s\langle g_s^2 GG \rangle\langle g_s \bar q \sigma Gq \rangle}{2880\pi^2}\omega ]e^{-\omega/T}d\omega \, ,
\\ && f_{\Xi_b,3,1,\rho\lambda}^{2} \Sigma_{\Xi_b,3,1,\rho\lambda} e^{-2 \bar \Lambda_{\Xi_b,3,1,\rho\lambda} / T}
\label{eq:Sc}
\\ \nonumber &=& \int_{2m_s}^{\omega_c} [ \frac{\langle g_s^2 GG \rangle}{4800\pi^4} \omega^7 - \frac{7m_s^2\langle g_s^2 GG \rangle}{2400\pi^4} \omega^5
\\ \nonumber && + \frac{7m_s\langle g_s^2 GG \rangle\langle \bar s s \rangle}{240\pi^2} \omega^3 ]e^{-\omega/T}d\omega \, .
\end{eqnarray}
Sum rules for other multiplets are listed in Appendix~\ref{app:sumrule}.

\section{Numerical Analyses}
\label{sec:numerical}

We use the following values for the $strange$ quark mass and various quark and gluon condensates~\cite{pdg,Yang:1993bp,Hwang:1994vp,Narison:2002pw,Gimenez:2005nt,Jamin:2002ev,Ioffe:2002be,Ovchinnikov:1988gk,colangelo}:
%
\begin{eqnarray}
\nonumber m_s(2~{\rm GeV}) &=& 95^{+9}_{-3} \mbox{ MeV} \, ,
\\ \nonumber \langle \bar qq \rangle &=& - (0.24 \pm 0.01)^3 \mbox{ GeV}^3 \, ,
\\ \nonumber \langle \bar ss \rangle &=& 0.8 \times \langle\bar qq \rangle \, ,
\\ \label{condensates} \langle g_s \bar q \sigma G q \rangle &=& M_0^2 \times \langle \bar qq \rangle\, ,
\\ \nonumber \langle g_s \bar s \sigma G s \rangle &=& M_0^2 \times \langle \bar ss \rangle\, ,
\\ \nonumber M_0^2 &=& 0.8 \mbox{ GeV}^2\, ,
\\ \nonumber \langle g_s^2GG\rangle &=& (0.48\pm 0.14) \mbox{ GeV}^4\, .
\end{eqnarray}
We also need the $bottom$ quark mass. In the present study we use the PDG value $m_b(m_b) = 4.18 ^{+0.04}_{-0.03}$~GeV~\cite{pdg} in the $\overline{\rm MS}$ scheme, instead of the pole mass $m_b = 4.78 \pm 0.06$~GeV~\cite{pdg2}. By~doing this, we intend to verify whether the masses extracted from the $[\mathbf{\bar 3}_F,3,1,\rho\lambda]$ multiplet can be consistent with the LHCb and CMS experiments~\cite{Aaij:2019amv,Sirunyan:2020gtz}.

There are two free parameters in Equations~(\ref{eq:ope})--(\ref{eq:Sc}), the threshold value $\omega_c$ and the Borel mass $T$, and~we use three criteria to constrain them: (a) The convergence of the Operator Product Expansion (OPE), (b) the Pole Contribution (PC), and (c) the mass dependence on $M_B$ and $s_0$. We refer to~\cite{Chen:2016phw,Mao:2017wbz} for detailed discussions, and here we just use the bottom baryon doublet $[\Xi_b(\mathbf{\bar 3}_F),3,1,\rho\lambda]$ as an example.

\begin{figure}[hbt]
\begin{center}
\scalebox{0.6}{\includegraphics{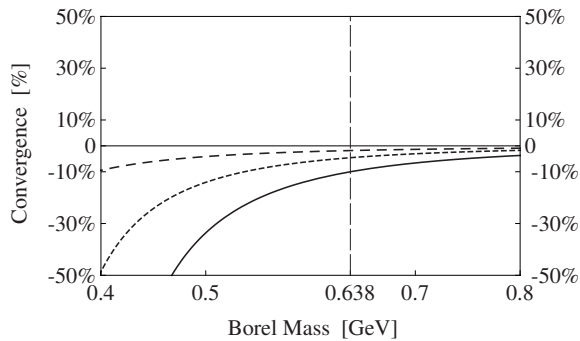}}
\caption{Variations of CVG$^{(\prime,\prime\prime)}$, defined in Equations~(\ref{eq:cvg1})--(\ref{eq:cvg3}), as functions of the Borel mass $T$, shown as the solid, short-dashed and long-dashed curves, respectively.}
\label{fig:CVG}
\end{center}
\end{figure}

The first criterion is the OPE convergence, which is the cornerstone of a reliable QCD sum rule analysis. In the present study we have calculated the sum rules at the leading order up to the eighth dimension, as shown in Equation~(\ref{eq:ope}); we have calculated the sum rules at the ${\mathcal O}(1/m_b)$ order up to the tenth dimension, as shown in Equations~(\ref{eq:Kc}) and (\ref{eq:Sc}). We investigate its convergence by requiring the high-order corrections of Equation~(\ref{eq:ope}) ($D=4+6+8$ terms) to be less than 10\%:
\begin{eqnarray}
{\rm CVG} \equiv { \Pi^{\rm D=4+6+8}_{\Xi_b,3,1,\rho\lambda}(\infty, T) \over \Pi_{\Xi_b,3,1,\rho\lambda}(\infty, T) } \leq 10 \% \, .
\label{eq:cvg1}
\end{eqnarray}
We show its variation with respect to the Borel mass $T$ in Figure~\ref{fig:CVG} using the solid curve when setting $\omega_c = 4.6$ GeV. We find that this condition is satisfied when $T>0.638$~GeV. Besides, it is also important to check the convergence of Equations~(\ref{eq:Kc}) and (\ref{eq:Sc}). To see this, we show:
\begin{eqnarray}
{\rm CVG}^\prime &\equiv& { K^{\rm D=4+6+8+10}_{\Xi_b,3,1,\rho\lambda}(\infty, T) \over K_{\Xi_b,3,1,\rho\lambda}(\infty, T) } \, ,
\label{eq:cvg2}
\\
{\rm CVG}^{\prime\prime} &\equiv& { \Sigma^{\rm D=4+6+8+10}_{\Xi_b,3,1,\rho\lambda}(\infty, T) \over \Sigma_{\Xi_b,3,1,\rho\lambda}(\infty, T) } \, ,
\label{eq:cvg3}
\end{eqnarray}
in Figure~\ref{fig:CVG} using the short-dashed and long-dashed curves. We find that their convergence is even~better.

\begin{figure*}[htbp]
\begin{center}
\scalebox{0.45}{\includegraphics{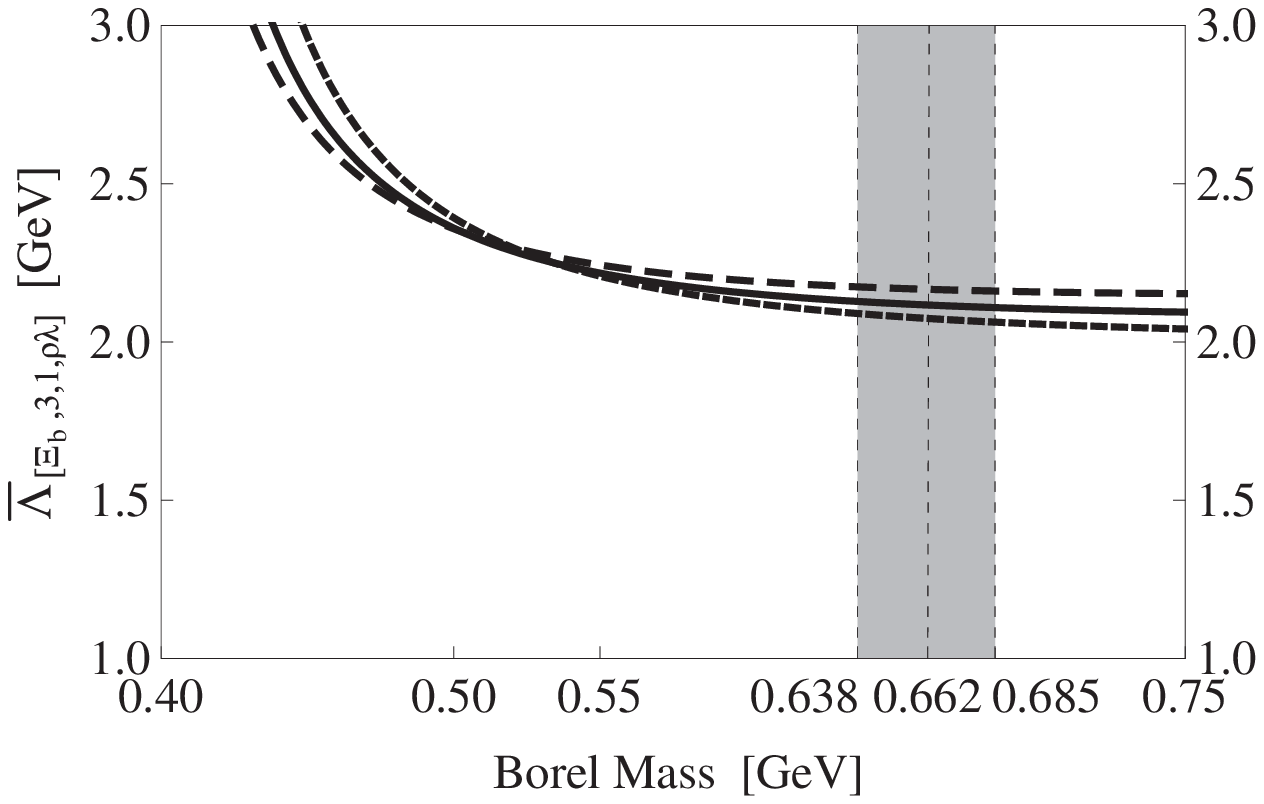}}~~~
\scalebox{0.45}{\includegraphics{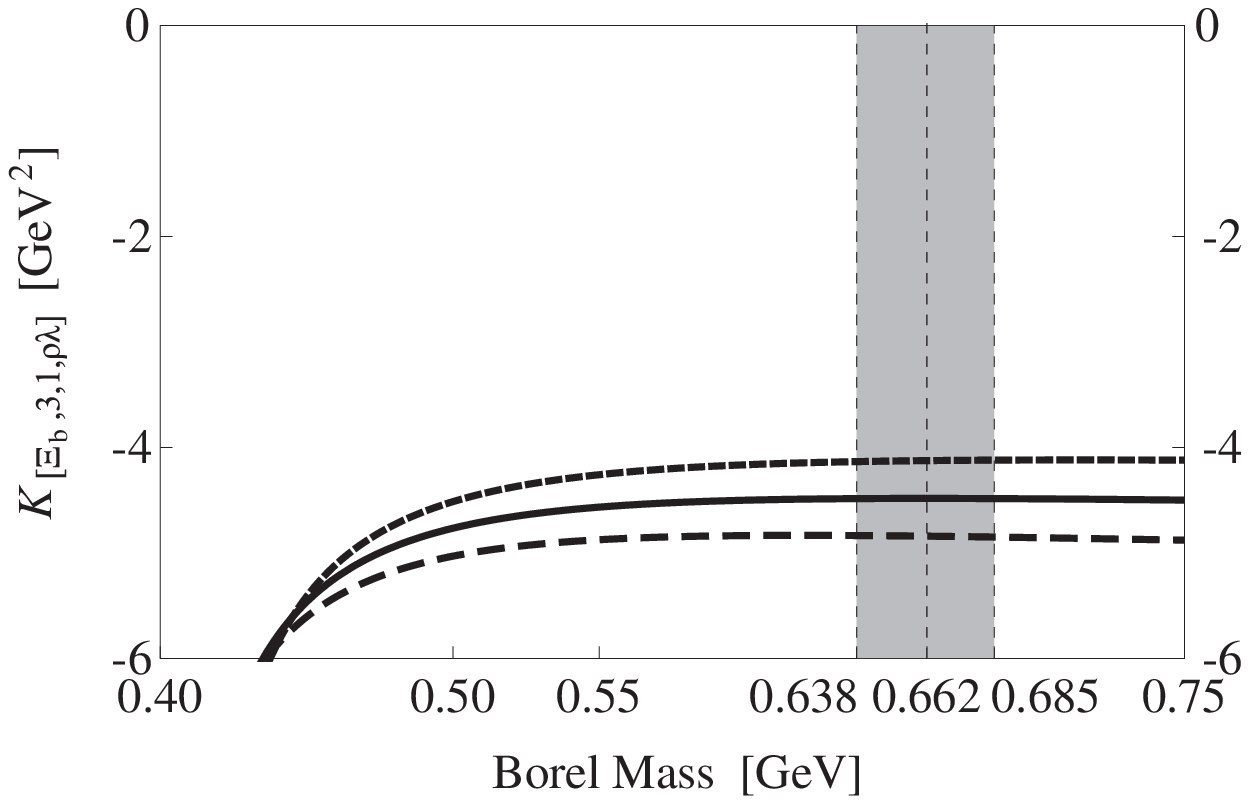}}~~~
\scalebox{0.45}{\includegraphics{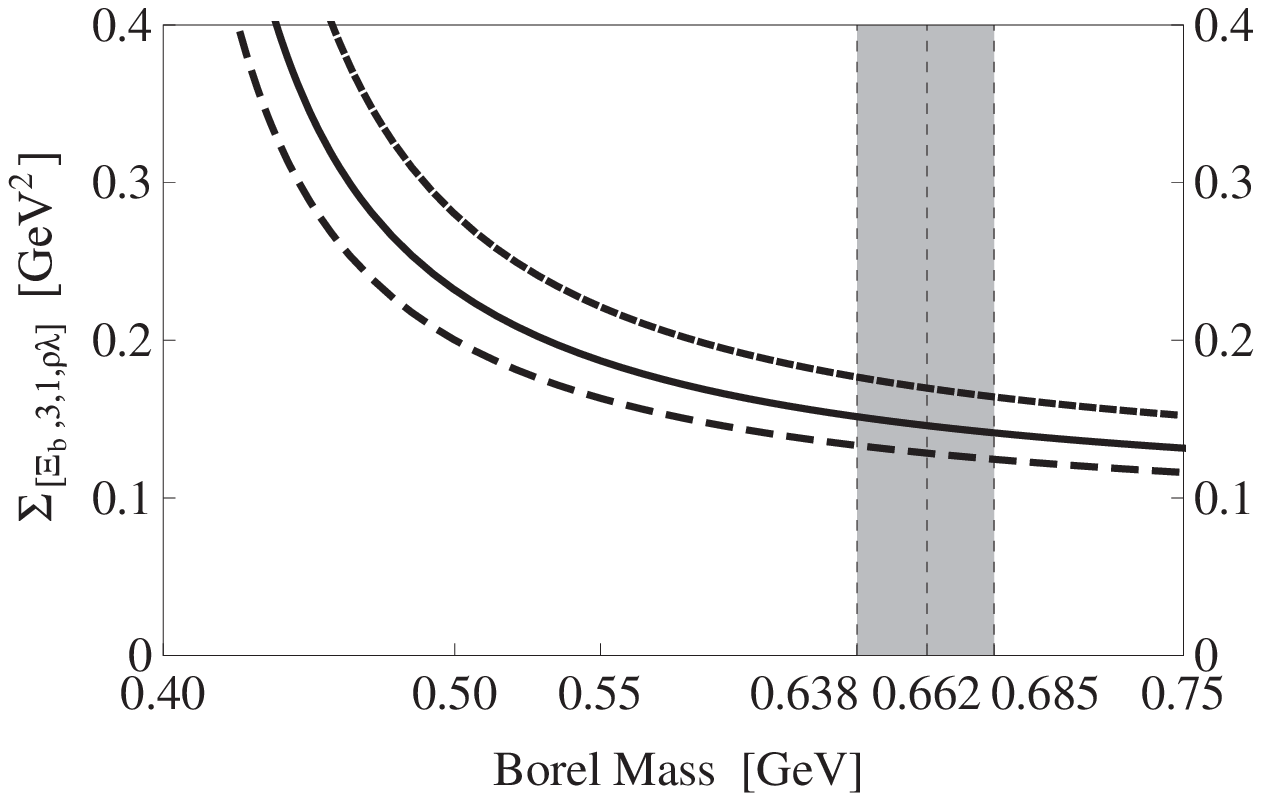}}
\caption{Variations of $\overline{\Lambda}_{\Xi_b,3,1,\rho\lambda}$ ({left}), $K_{\Xi_b,3,1,\rho\lambda}$ ({middle}), and $\Sigma_{\Xi_b,3,1,\rho\lambda}$ ({right}) as functions of the Borel mass $T$, where the short-dashed, solid, and long-dashed curves are obtained by fixing $\omega_c = 4.4$, $4.6$, and $4.8$~GeV, respectively. These figures are plotted for the bottom baryon doublet $[\Xi_b(\mathbf{\bar 3}_F),3,1,\rho\lambda]$.
}
\label{fig:sumrule}
\end{center}
\end{figure*}

The second criterion is requires the PC to be larger than 10\%:
%
\begin{equation}
\label{eq_pole}
\mbox{PC} \equiv \frac{ \Pi_{\Xi_b,3,1,\rho\lambda}(\omega_c, T) }{  \Pi_{\Xi_b,3,1,\rho\lambda}(\infty, T) } \geq 10\% \, .
\end{equation}
%
This condition is satisfied when $T< 0.685$~GeV. Altogether, we choose $\omega_c = 4.6$ GeV and extract the Borel window to be $0.638$~GeV$< T < 0.685$~GeV, from which we obtain:
\begin{eqnarray}
\nonumber \bar \Lambda_{\Xi_b(\mathbf{\bar 3}_F),3,1,\rho\lambda} &=& 2.117 \mbox{ GeV} \, , \,
\\ K_{\Xi_b(\mathbf{\bar 3}_F),3,1,\rho\lambda} &=& -4.483 \mbox{ GeV}^2 \, , \,
\\ \nonumber \Sigma_{\Xi_b(\mathbf{\bar 3}_F),3,1,\rho\lambda} &=& 0.018 \mbox{ GeV}^{2} \, .
\end{eqnarray}
Their variations are shown in Figure~\ref{fig:sumrule} as functions of the Borel mass $T$, where their $T$ dependence is weak and acceptable inside the Borel window $0.638$~GeV$< T < 0.685$~GeV.

\begin{figure*}[htbp]
\begin{center}
\begin{tabular}{c}
\scalebox{0.6}{\includegraphics{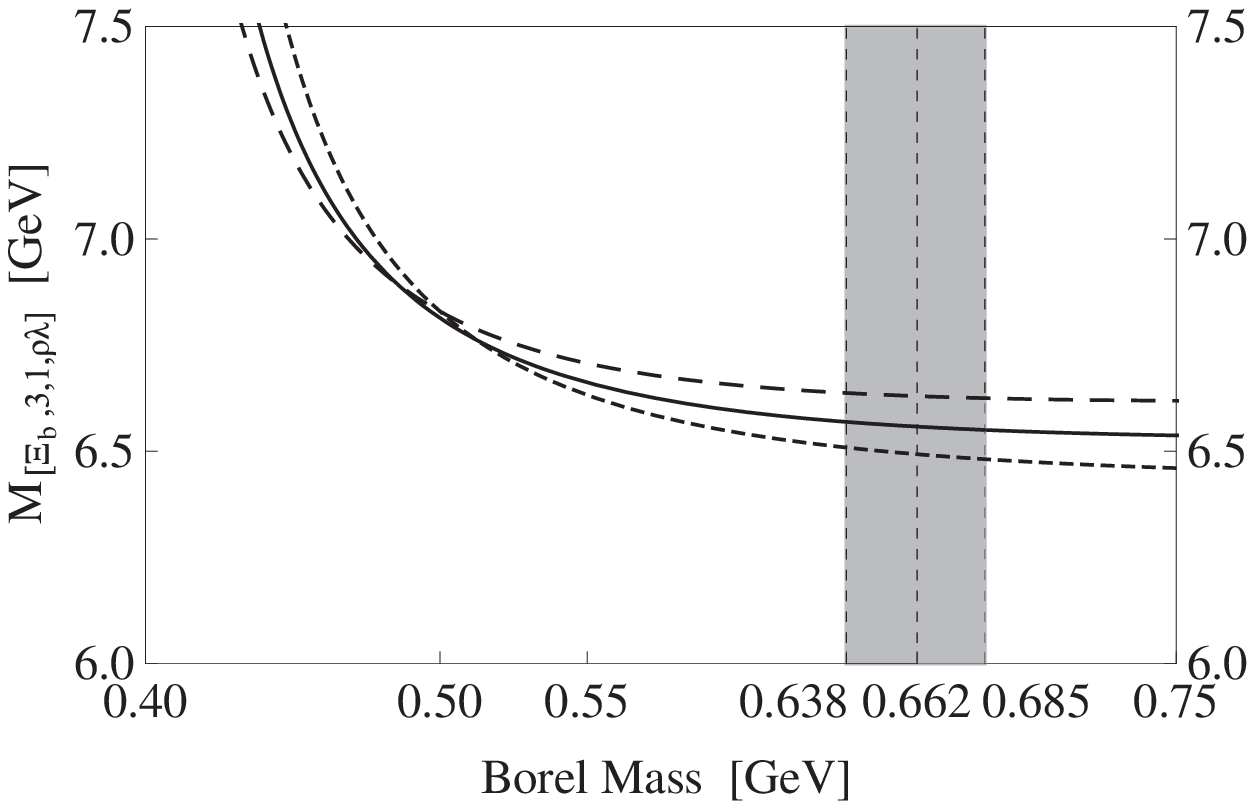}}~~~~~
\scalebox{0.598}{\includegraphics{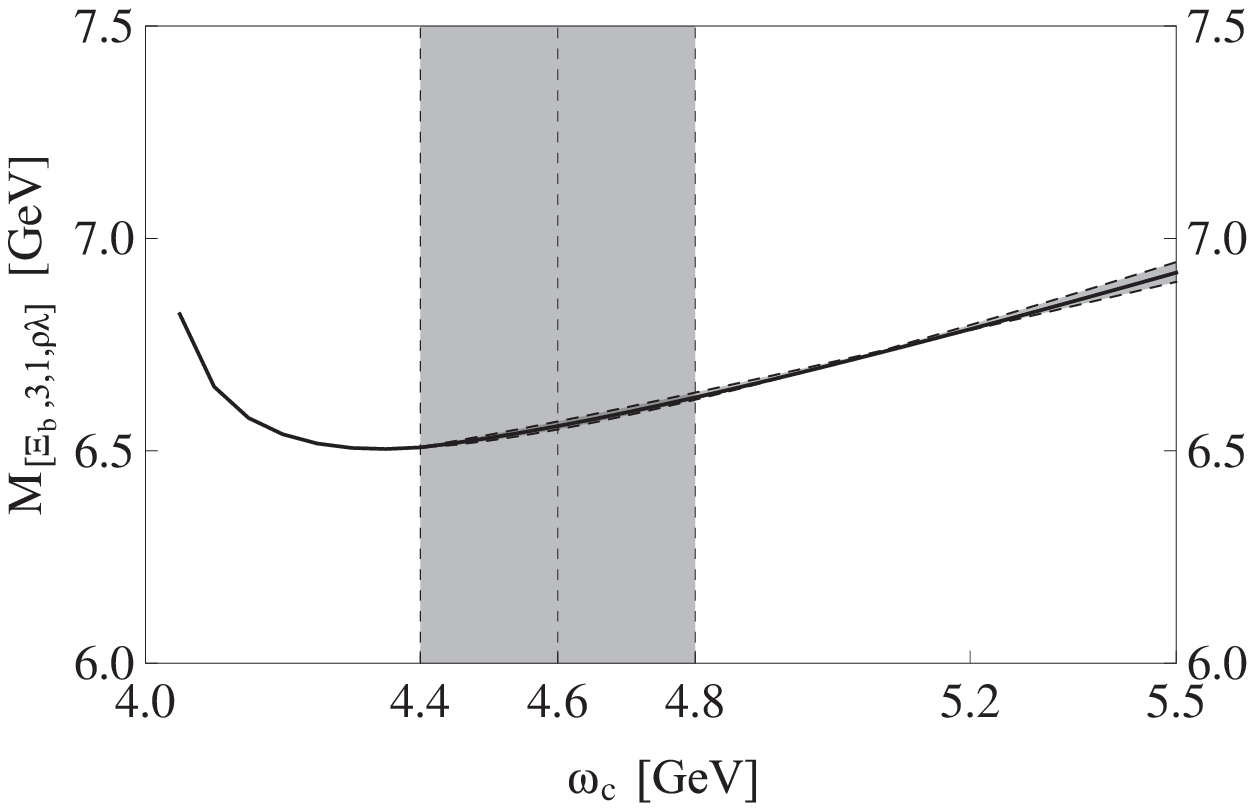}}
\end{tabular}
\caption{The variation of $m_{\Xi_b(5/2^+)}$ as a function of the Borel mass $T$ ({left}) and the threshold value $\omega_c$ ({right}). In the left panel, the short-dashed, solid, and long-dashed curves are obtained by setting $\omega_c = 4.4$, 4.6, and 4.8~GeV, respectively. In the right panel the shady band is obtained by varying $T$ within the Borel window. These figures are plotted for the bottom baryon doublet $[\Xi_b(\mathbf{\bar 3}_F),3,1,\rho\lambda]$.
}
\label{fig:mass3111us}
\end{center}
\end{figure*}

Then we use Equations~(\ref{eq:mass}) and (\ref{eq:masscorrection}) to further obtain:
\begin{eqnarray}
\nonumber m_{\Xi_b(5/2^+)} &=& 6.56 \mbox{ GeV} \, , \,
\\ m_{\Xi_b(7/2^+)} &=& 6.57 \mbox{ GeV} \, , \,
\\ \nonumber \Delta m_{[\Xi_b(\mathbf{\bar 3}_F),3,1,\rho\lambda]} &=& 12.2 \mbox{ MeV} \, ,
\end{eqnarray}
where $m_{\Xi_b(5/2^+)}$ and $m_{\Xi_b(7/2^+)}$ are the masses of the $\Xi_b(5/2^+)$ and $\Xi_b(7/2^+)$ belonging to the $[\Xi_b(\mathbf{\bar 3}_F),3,1,\rho\lambda]$ multiplet, with $\Delta m_{[\Xi_b(\mathbf{\bar 3}_F),3,1,\rho\lambda]}$ their mass splitting. The variation of $m_{\Xi_b(5/2^+)}$ is shown in the left panel of Figure~\ref{fig:mass3111us} as a function of the Borel mass $T$, where its $T$ dependence is also weak and acceptable inside the Borel window $0.638$~GeV$< T < 0.685$~GeV.

We change the threshold value $\omega_c$ and redo the above procedures. The variation of $m_{\Xi_b(5/2^+)}$ is shown in the right panel of Figure~\ref{fig:mass3111us} as a function of the threshold value $\omega_c$. We find that there exist non-vanishing Borel windows as long as $\omega_c \geq 4.4$~GeV, and the $\omega_c$ dependence is weak and acceptable in the region $4.4$~GeV$<\omega_c<4.8$~GeV. This is just the working region for $\omega_c$, where both the mass $m_{\Xi_b(5/2^+)}$ as well as its uncertainty can be evaluated reliably.

Hence, we fix our working regions to be $4.4$~GeV$<\omega_c<4.8$~GeV and $0.638$~GeV$< T < 0.685$~GeV, and obtain:
\begin{eqnarray}
\nonumber m_{\Xi_b(5/2^+)} &=& 6.56^{+0.12}_{-0.10} \mbox{ GeV} \, , \,
\\ m_{\Xi_b(7/2^+)} &=& 6.57^{+0.12}_{-0.10} \mbox{ GeV} \, , \,
\\ \nonumber \Delta m_{[\Xi_b(\mathbf{\bar 3}_F),3,1,\rho\lambda]} &=& 12.2^{+6.3}_{-4.8} \mbox{ MeV} \, ,
\end{eqnarray}
where the central values correspond to $\omega_c = 4.6$~GeV and $T=0.662$~GeV, and the uncertainties are due to the threshold value $\omega_c$, the Borel mass $T$, the $strange$ and $bottom$ quark masses, and various quark and gluon condensates.

\begin{figure*}[htbp]
\begin{center}
\begin{tabular}{c}
\scalebox{0.6}{\includegraphics{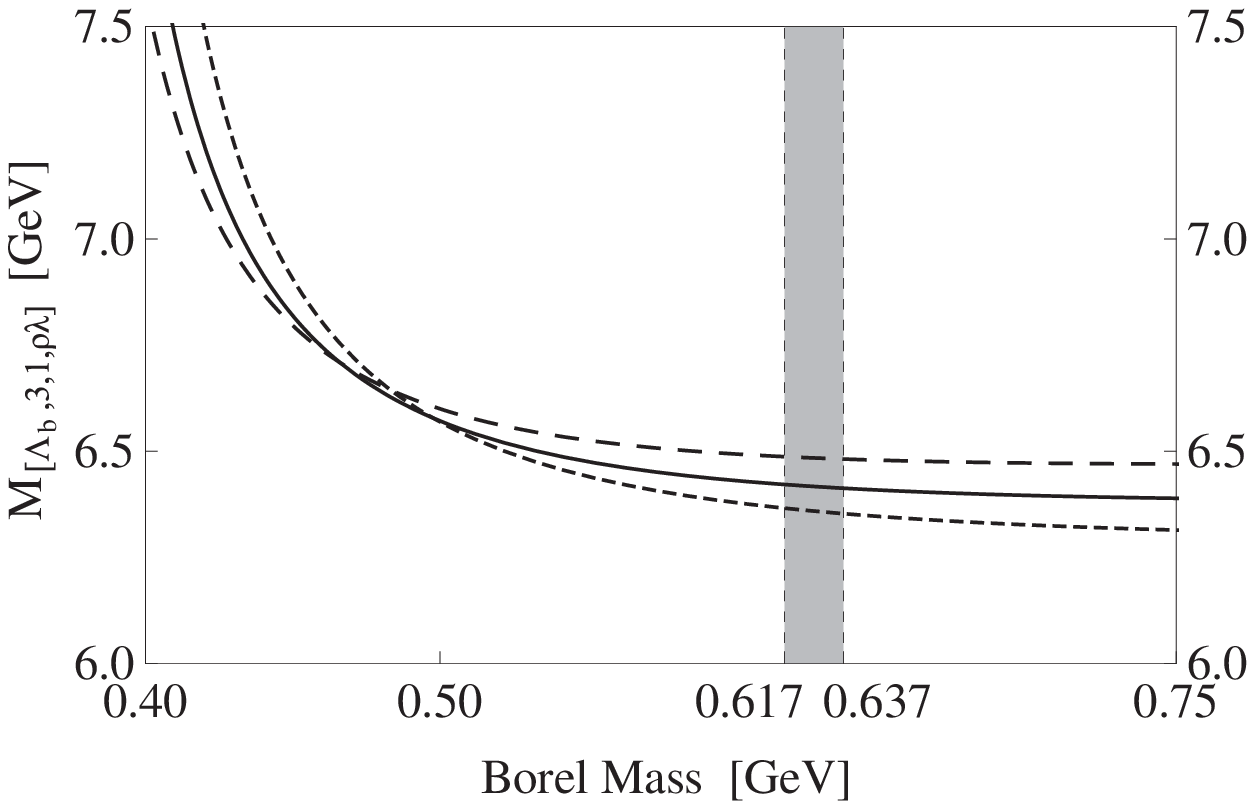}}~~~~~
\scalebox{0.598}{\includegraphics{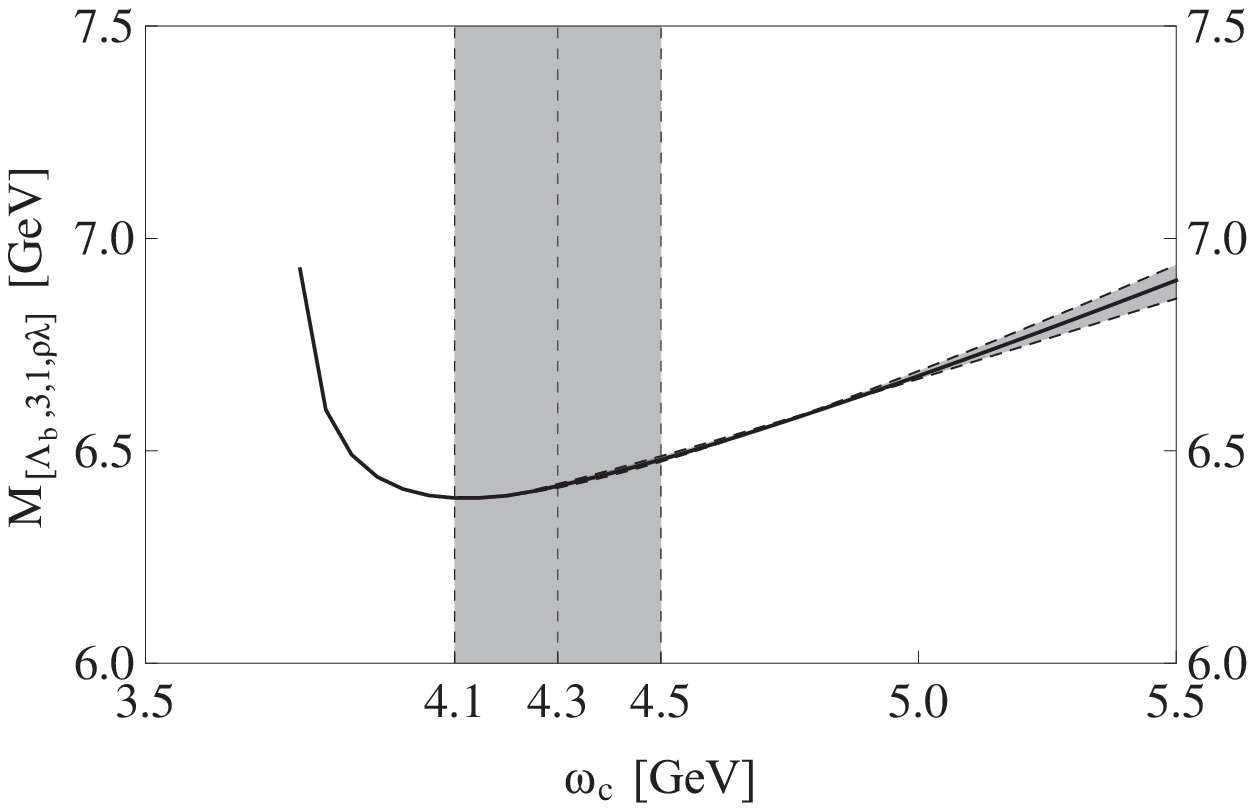}}
\end{tabular}
\caption{The variation of $m_{\Lambda_b(5/2^+)}$ as a function of the Borel mass $T$ ({left}) and the threshold value $\omega_c$ ({right}). In the left panel, the short-dashed, solid ,and long-dashed curves are obtained by setting $\omega_c = 4.1$, 4.3, and 4.5~GeV, respectively. In the right panel the shady band is obtained by varying $T$ within the Borel window. These figures are plotted for the bottom baryon doublet $[\Lambda_b(\mathbf{\bar 3}_F),3,1,\rho\lambda]$.
}
\label{fig:mass3111ud}
\end{center}
\end{figure*}

Following the same procedures, we study the bottom baryon doublet $[\Lambda_b(\mathbf{\bar 3}_F),3,1,\rho\lambda]$, which contains the $\Lambda_b(5/2^+)$ and $\Lambda_b(7/2^+)$. They are the partner states of the $\Xi_b(5/2^+)$ and $\Xi_b(7/2^+)$ belonging to the $[\Xi_b(\mathbf{\bar 3}_F),3,1,\rho\lambda]$ multiplet, and their masses are extracted to be:
\begin{eqnarray}
\nonumber m_{\Lambda_b(5/2^+)} &=& 6.42^{+0.15}_{-0.11} \mbox{ GeV} \, , \,
\\ m_{\Lambda_b(7/2^+)} &=& 6.43^{+0.15}_{-0.11} \mbox{ GeV} \, , \,
\\ \nonumber \Delta m_{[\Lambda_b(\mathbf{\bar 3}_F),3,1,\rho\lambda]} &=& 14.6^{+8.8}_{-6.2} \mbox{ MeV} \, .
\end{eqnarray}
For completeness, we show the variation of $m_{\Lambda_b(5/2^+)}$ in Figure~\ref{fig:mass3111ud} as a function of the Borel mass $T$ (left) and the threshold value $\omega_c$ (right).

\begin{figure*}[htbp]
\begin{center}
\begin{tabular}{c}
\scalebox{0.6}{\includegraphics{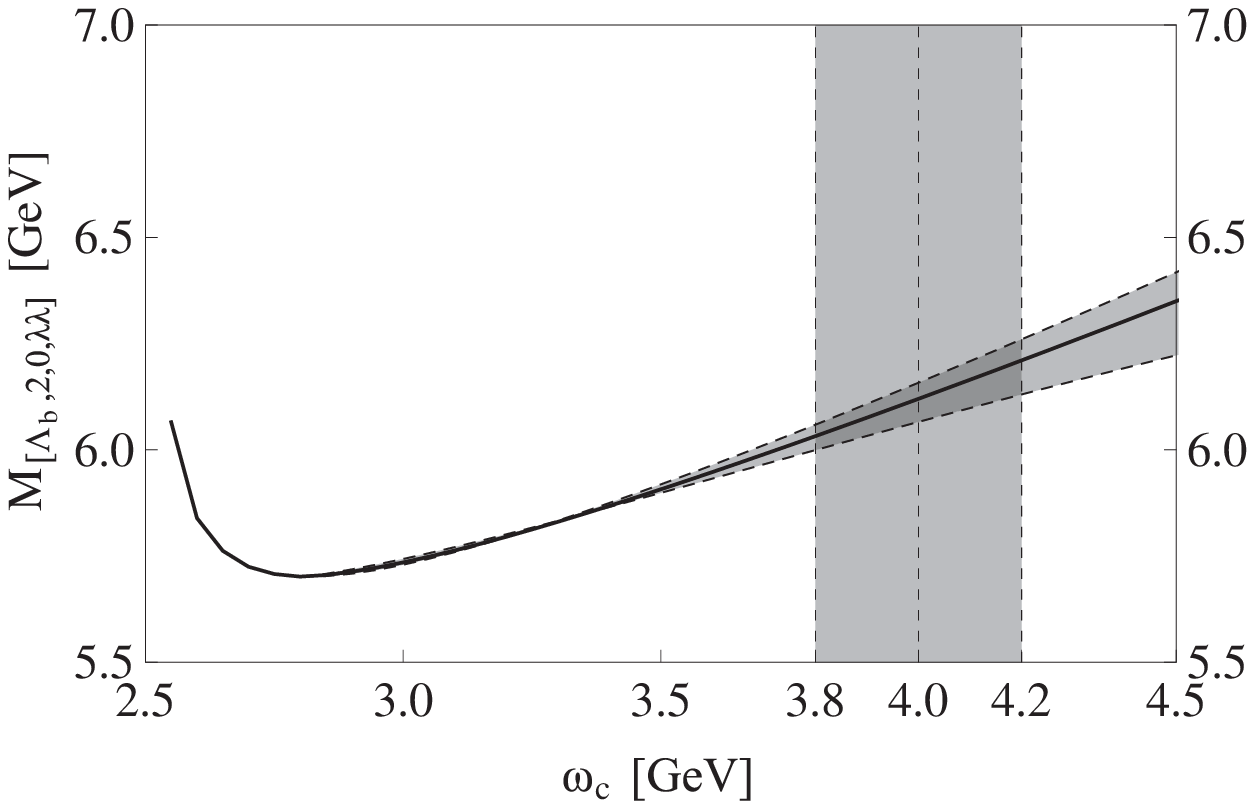}}~~~~~
\scalebox{0.598}{\includegraphics{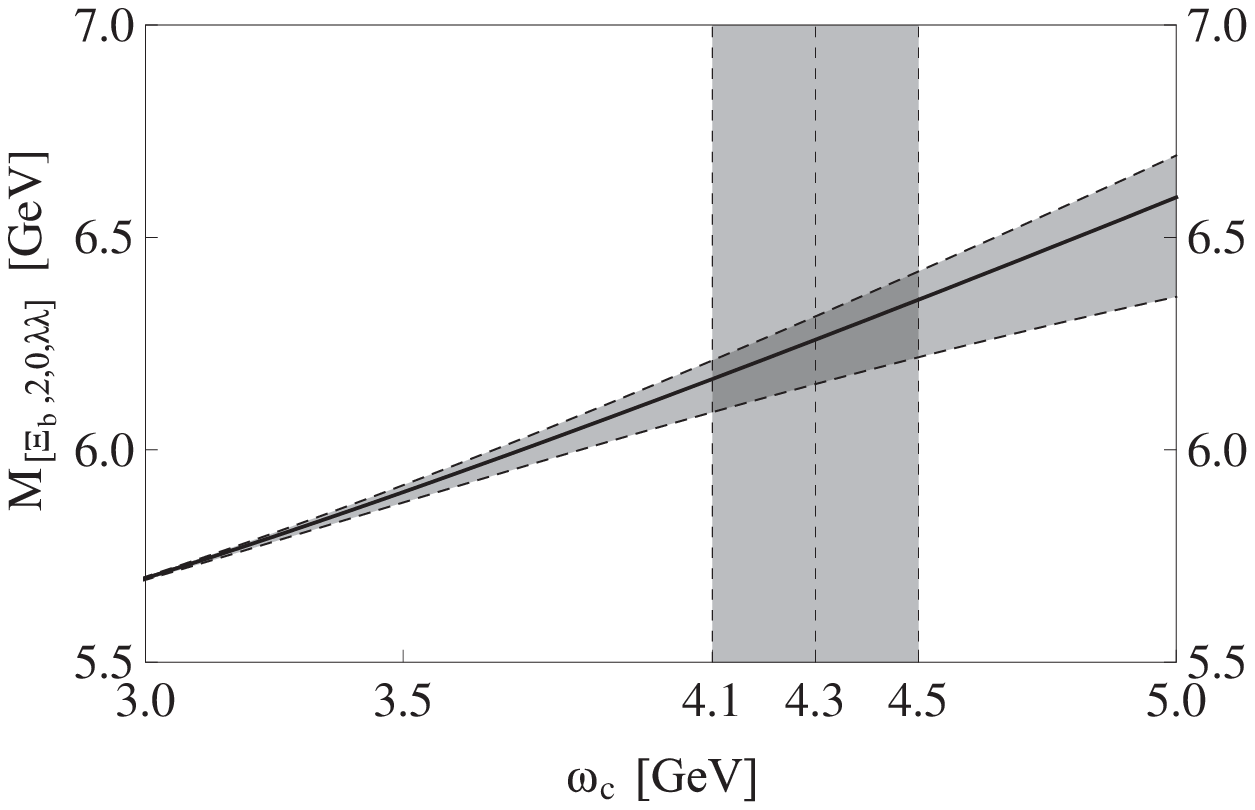}}
\\
\scalebox{0.6}{\includegraphics{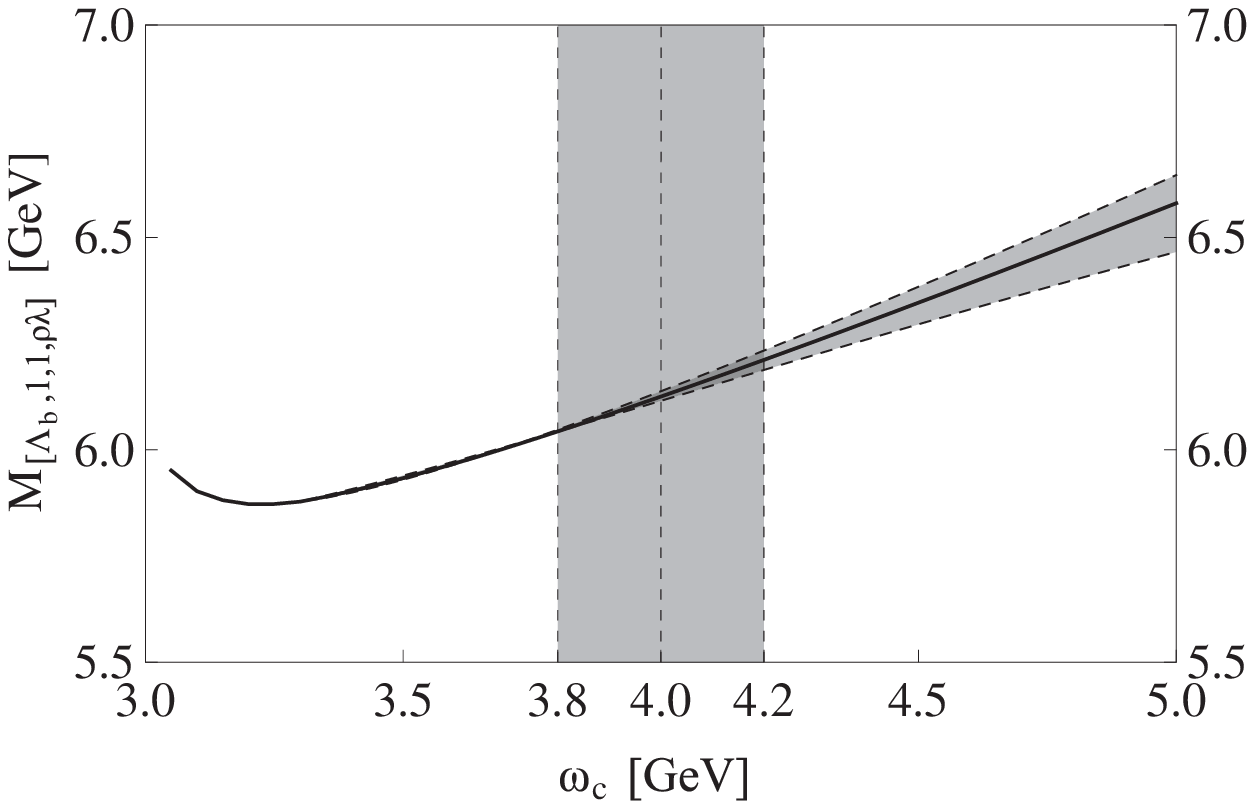}}~~~~~
\scalebox{0.598}{\includegraphics{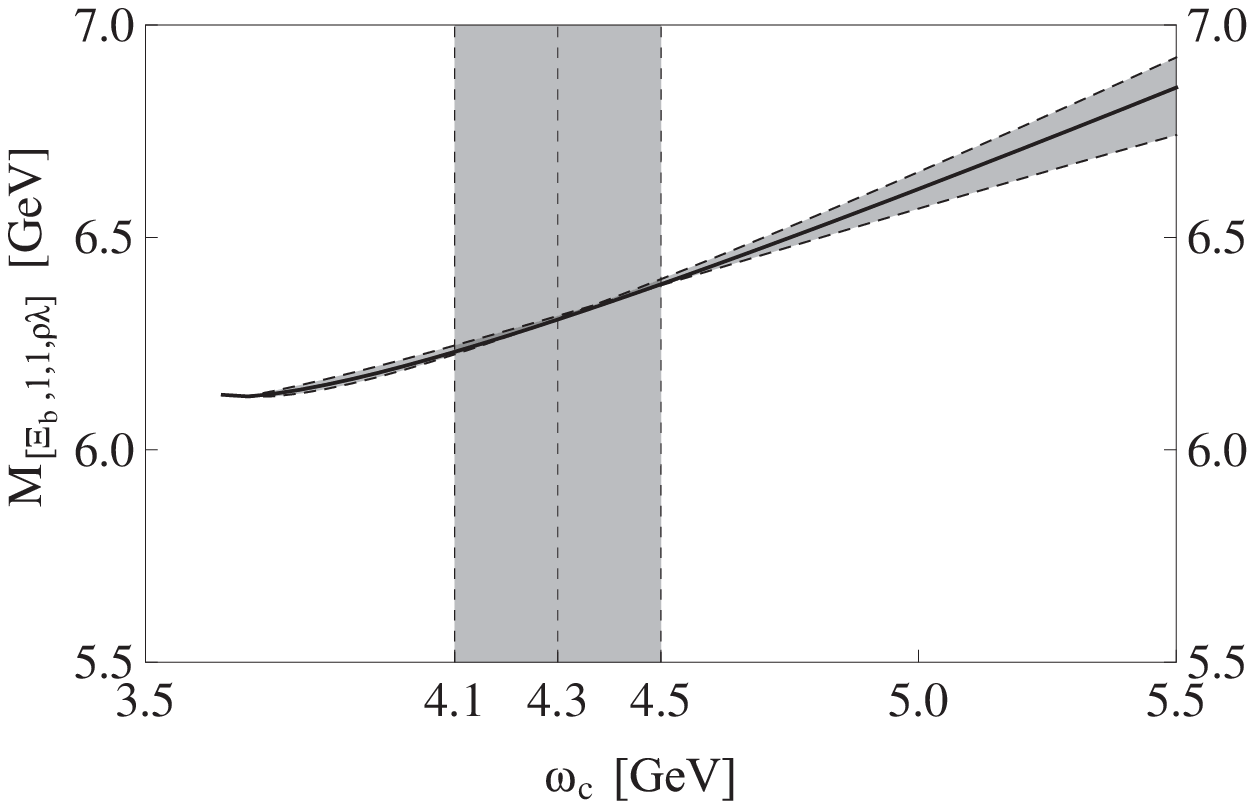}}
\\
\scalebox{0.6}{\includegraphics{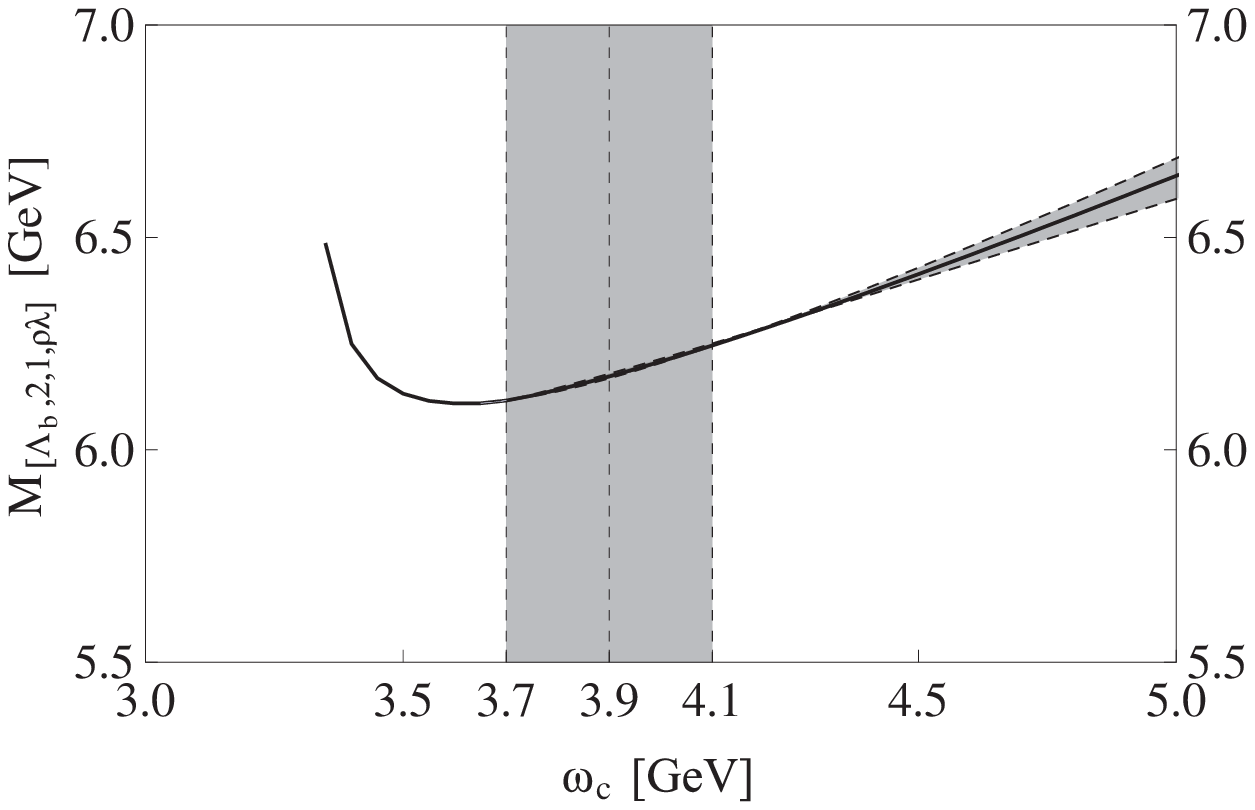}}~~~~~
\scalebox{0.598}{\includegraphics{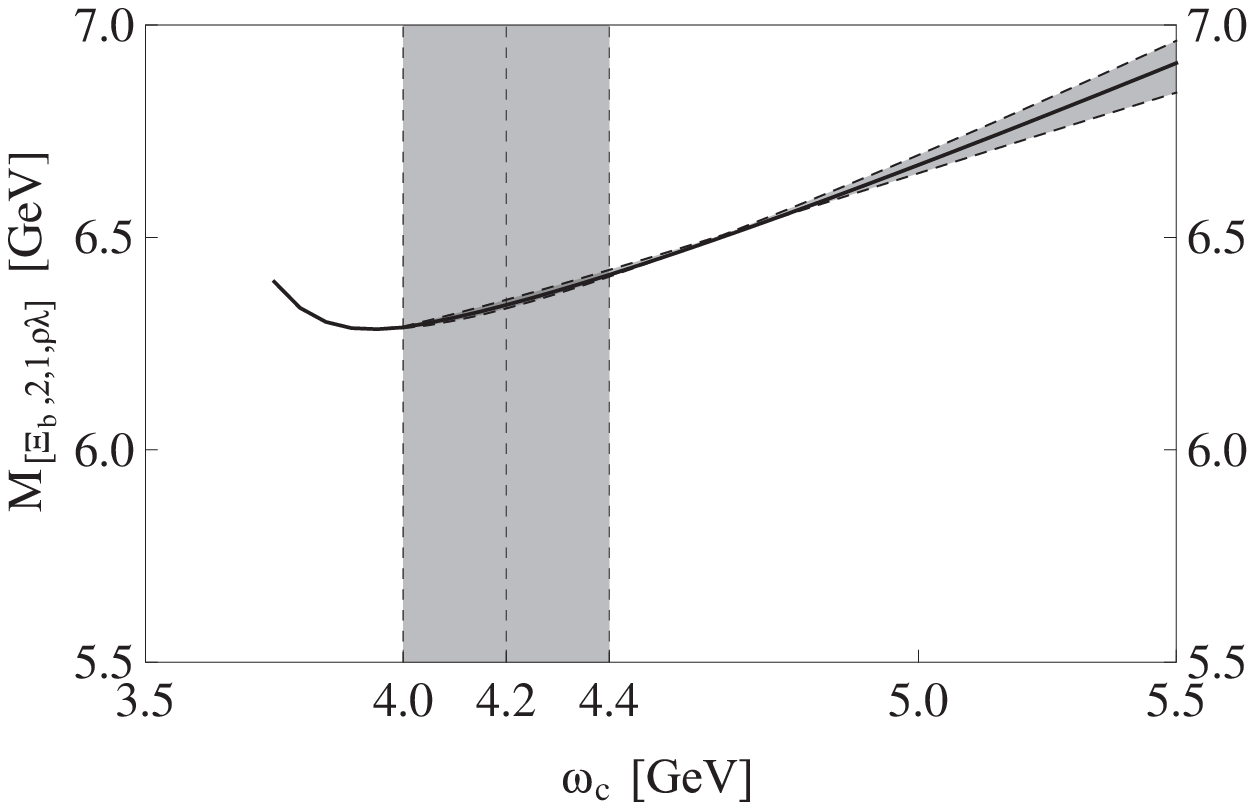}}
\end{tabular}
\caption{Masses as functions of the Borel mass $T$, extracted from the bottom baryon doublets: $[\Lambda_b(\mathbf{\bar 3}_F),2,0,\lambda\lambda]$ ({upper-left}), $[\Xi_b(\mathbf{\bar 3}_F),2,0,\lambda\lambda]$ ({upper-right}), $[\Lambda_b(\mathbf{\bar 3}_F),1,1,\rho\lambda]$ ({middle-left}), $[\Xi_b(\mathbf{\bar 3}_F),1,1,\rho\lambda]$ ({middle-right}), $[\Lambda_b(\mathbf{\bar 3}_F),3,1,\rho\lambda]$ ({lower-left}), and $[\Xi_b(\mathbf{\bar 3}_F),3,1,\rho\lambda]$ ({lower-right}). The~shady bands are obtained by varying $T$ within the Borel window.
}
\label{fig:mass}
\end{center}
\end{figure*}

Similarly, we study the $[\mathbf{\bar 3}_F,2,0,\rho\rho]$, $[\mathbf{\bar 3}_F,2,0,\lambda\lambda]$, $[\mathbf{\bar 3}_F,1,1,\rho\lambda]$, and $[\mathbf{\bar 3}_F,2,1,\rho\lambda]$ multiplets. The~latter three lead to reasonable sum rule results, and the extracted masses are shown in Figure~\ref{fig:mass} as functions of the threshold value $\omega_c$. We summarize all the above results in Table~\ref{tab:results}, which will be discussed in the next section.

\section{Summary and Discussions}
\label{sec:summary}

\begin{table*}[hbt]
\begin{center}
\renewcommand{\arraystretch}{1.6}
\caption{Masses of the $D$-wave bottom baryons belonging to the $SU(3)$ flavor $\mathbf{\bar 3}_F$ representation, obtained using the bottom baryon multiplets $[\mathbf{\bar 3}_F, 2, 0, \lambda\lambda]$, $[\mathbf{\bar 3}_F, 1, 1, \rho\lambda]$, $[\mathbf{\bar 3}_F, 2, 1, \rho\lambda]$, and $[\mathbf{\bar 3}_F, 3, 1, \rho\lambda]$. There still exists one doublet, the $[\mathbf{\bar 3}_F, 2, 0, \rho\rho]$ doublet, but its sum rule results are a bit strange (the Borel windows become larger as the threshold value $\omega_c$ decreases), so we do not list it here. In the third and fourth columns we list the two {QCD} sum rule parameters, the threshold value $\omega_c$ and the Borel mass $T$; in the fifth, sixth, and seventh columns we list the QCD sum rule results, $\overline{\Lambda}_{F,j_l,s_l,\rho/\lambda}$ at the leading order as well as $K_{F,j_l,s_l,\rho/\lambda}$ and $\Sigma_{F,j_l,s_l,\rho/\lambda}$ at the ${\mathcal O}(1/m_b)$ order; in the ninth, tenth, and eleventh columns we list the masses $m_{j,P,F,j_l,s_l,\rho/\lambda}$, decay constants $f_{F,j_l,s_l,\rho\rho/\lambda\lambda/\rho\lambda}$, and mass splittings $\Delta M_{F,j_l,s_l,\rho\rho/\lambda\lambda/\rho\lambda} = m_{j_l+1/2,P,F,j_l,s_l,\rho/\lambda} - m_{j_l-1/2,P,F,j_l,s_l,\rho/\lambda}$ evaluated within the QCD sum rule method.
}
\begin{tabular}{c | c | c | c | c c c | c | c | c | c}
\hline\hline
\multirow{2}{*}{Multiplets} & \multirow{2}{*}{~B~} & $\omega_c$ & ~~~Working region~~~ & ~~~~~~~$\overline{\Lambda}$~~~~~~~ & ~~~~~$K$~~~~~ & ~~~~~$\Sigma$~~~~~ & ~Baryons~ & ~~~~Mass~~~~ & ~~~~~~~$f$~~~~~~~ & $\Delta M$
\\ & & (GeV) & (GeV) & (GeV) & (GeV$^2$) & (GeV$^2$) & ($j^P$) & (GeV) & (GeV$^{5}$) & (MeV)
\\ \hline\hline
\multirow{4}{*}{$[\mathbf{\bar 3}_F, 2, 0, \lambda\lambda]$} & \multirow{2}{*}{$\Lambda_b$} & \multirow{2}{*}{4.0} & \multirow{2}{*}{$0.412< T < 0.632$} & \multirow{2}{*}{$1.680^{+0.073}_{-0.082}$} & \multirow{2}{*}{$-4.398$} & \multirow{2}{*}{$0.0112$} & $\Lambda_b(3/2^+)$ & $6.12^{+0.10}_{-0.11}$ & $0.114^{+0.029}_{-0.026}$ &
\multirow{2}{*}{$5.4^{+2.3}_{-1.9}$}
\\ \cline{8-10}
& & & & & & & $\Lambda_b(5/2^+)$ & $6.13^{+0.10}_{-0.11}$ & $0.048^{+0.012}_{-0.011}$ &
\\ \cline{2-11}
& \multirow{2}{*}{$\Xi_b$} & \multirow{2}{*}{4.3} & \multirow{2}{*}{$0.414< T < 0.678$} & \multirow{2}{*}{$1.792^{+0.079}_{-0.109}$} & \multirow{2}{*}{$-4.846$} & \multirow{2}{*}{$0.0095$} & $\Xi_b(3/2^+)$ & $6.26^{+0.11}_{-0.14}$ & $0.158^{+0.038}_{-0.041}$ & \multirow{2}{*}{$4.5^{+1.9}_{-1.5}$}
\\ \cline{8-10}
& & & & & & & $\Xi_b(5/2^+)$ & $6.26^{+0.11}_{-0.14}$ & $0.067^{+0.016}_{-0.017}$ &
\\ \hline
\multirow{4}{*}{$[\mathbf{\bar 3}_F, 1, 1, \rho\lambda]$} & \multirow{2}{*}{$\Lambda_b$} & \multirow{2}{*}{4.0} & \multirow{2}{*}{$0.483< T < 0.621$} & \multirow{2}{*}{$1.732^{+0.071}_{-0.069}$} & \multirow{2}{*}{$-3.555$} & \multirow{2}{*}{$-0.0050$} & $\Lambda_b(1/2^+)$ & $6.13^{+0.10}_{-0.09}$ & $0.159^{+0.040}_{-0.034}$ & \multirow{2}{*}{$-1.5^{+0.6}_{-0.5}$}
\\ \cline{8-10}
& & & & & & & $\Lambda_b(3/2^+)$ & $6.13^{+0.10}_{-0.09}$ & $0.075^{+0.019}_{-0.016}$ &
\\ \cline{2-11}
& \multirow{2}{*}{$\Xi_b$} & \multirow{2}{*}{4.3} & \multirow{2}{*}{$0.532< T < 0.661$} & \multirow{2}{*}{$1.902^{+0.064}_{-0.060}$} & \multirow{2}{*}{$-3.732$} & \multirow{2}{*}{$-0.0053$} & $\Xi_b(1/2^+)$ & $6.31^{+0.09}_{-0.09}$ & $0.235^{+0.054}_{-0.046}$ & \multirow{2}{*}{$-1.5^{+0.6}_{-0.6}$}
\\ \cline{8-10}
& & & & & & & $\Xi_b(3/2^+)$ & $6.31^{+0.09}_{-0.09}$ & $0.111^{+0.026}_{-0.022}$ &
\\ \hline
\multirow{4}{*}{$[\mathbf{\bar 3}_F, 2, 1, \rho\lambda]$} & \multirow{2}{*}{$\Lambda_b$} & \multirow{2}{*}{3.9} & \multirow{2}{*}{$0.543< T < 0.585$} & \multirow{2}{*}{$1.772^{+0.096}_{-0.076}$} & \multirow{2}{*}{$-3.819$} & \multirow{2}{*}{$0.0283$} & $\Lambda_b(3/2^+)$ & $6.17^{+0.12}_{-0.10}$ & $0.219^{+0.059}_{-0.048}$ & \multirow{2}{*}{$13.5^{+7.5}_{-5.6}$}
\\ \cline{8-10}
& & & & & & & $\Lambda_b(5/2^+)$ & $6.19^{+0.12}_{-0.10}$ & $0.079^{+0.021}_{-0.017}$ &
\\ \cline{2-11}
& \multirow{2}{*}{$\Xi_b$} & \multirow{2}{*}{4.2} & \multirow{2}{*}{$0.576< T < 0.626$} & \multirow{2}{*}{$1.938^{+0.078}_{-0.063}$} & \multirow{2}{*}{$-3.853$} & \multirow{2}{*}{$0.0247$} & $\Xi_b(3/2^+)$ & $6.34^{+0.10}_{-0.09}$ & $0.327^{+0.078}_{-0.065}$ & \multirow{2}{*}{$11.8^{+5.8}_{-4.6}$}
\\ \cline{8-10}
& & & & & & & $\Xi_b(5/2^+)$ & $6.35^{+0.11}_{-0.09}$ & $0.118^{+0.028}_{-0.023}$ &
\\ \hline
\multirow{4}{*}{$[\mathbf{\bar 3}_F, 3, 1, \rho\lambda]$} & \multirow{2}{*}{$\Lambda_b$} & \multirow{2}{*}{4.3} & \multirow{2}{*}{$0.617< T < 0.637$} & \multirow{2}{*}{$1.986^{+0.122}_{-0.089}$} & \multirow{2}{*}{$-4.336$} & \multirow{2}{*}{$0.0218$} & $\Lambda_b(5/2^+)$ & $6.42^{+0.15}_{-0.11}$ & $0.388^{+0.096}_{-0.078}$ & \multirow{2}{*}{$14.6^{+8.8}_{-6.2}$}
\\ \cline{8-10}
& & & & & & & $\Lambda_b(7/2^+)$ & $6.43^{+0.15}_{-0.11}$ & $0.143^{+0.035}_{-0.029}$ &
\\ \cline{2-11}
& \multirow{2}{*}{$\Xi_b$} & \multirow{2}{*}{4.6} & \multirow{2}{*}{$0.638< T < 0.685$} & \multirow{2}{*}{$2.117^{+0.096}_{-0.076}$} & \multirow{2}{*}{$-4.483$} & \multirow{2}{*}{$0.0182$} & $\Xi_b(5/2^+)$ & $6.56^{+0.12}_{-0.10}$ & $0.543^{+0.120}_{-0.101}$ & \multirow{2}{*}{$12.2^{+6.3}_{-4.8}$}
\\ \cline{8-10}
& & & & & & & $\Xi_b(7/2^+)$ & $6.57^{+0.12}_{-0.10}$ & $0.200^{+0.044}_{-0.037}$ &
\\ \hline \hline
\end{tabular}
\label{tab:results}
\end{center}
\end{table*}

In this paper we apply the method of QCD sum rules within the heavy quark effective theory to study $D$-wave bottom baryons of the $SU(3)$ flavor $\mathbf{\bar 3}_F$. We investigate five bottom baryon doublets, $[\mathbf{\bar 3}_F, 2, 0, \rho\rho]$, $[\mathbf{\bar 3}_F, 2, 0, \lambda\lambda]$, $[\mathbf{\bar 3}_F, 1, 1, \rho\lambda]$, $[\mathbf{\bar 3}_F, 2, 1, \rho\lambda]$, and $[\mathbf{\bar 3}_F, 3, 1, \rho\lambda]$. Their masses are calculated up to the order $\mathcal{O}(1/m_b)$, and the results are summarized in Table~\ref{tab:results}, including the masses $m_{j,P,F,j_l,s_l,\rho/\lambda}$, mass splittings $\Delta M_{F,j_l,s_l,\rho\rho/\lambda\lambda/\rho\lambda}$, and decay constants $f_{F,j_l,s_l,\rho\rho/\lambda\lambda/\rho\lambda}$ evaluated in the present study. Before discussing these results, we note that there is a considerable uncertainty in our results for the absolute value of the mass, because it depends significantly on the bottom quark mass, as shown in Equation~(\ref{eq:mass}) however, the mass difference within the same doublet does not depend much on the bottom quark mass, so it is produced quite well with much less (theoretical) uncertainty and gives more useful information.

Based on Table~\ref{tab:results}, we conclude the present study:
\begin{itemize}

\item The masses of $\Lambda_b(3/2^+)$ and $\Lambda_b(5/2^+)$ calculated using the $[\mathbf{\bar 3}_F, 2, 0, \lambda\lambda]$ multiplet are:
\begin{eqnarray}
\nonumber m_{\Lambda_b(3/2^+)} &=& 6.12^{+0.10}_{-0.11} \mbox{ GeV} \, , \,
\\ m_{\Lambda_b(5/2^+)} &=& 6.13^{+0.10}_{-0.11} \mbox{ GeV} \, , \,
\\ \nonumber \Delta m_{[\Lambda_b(\mathbf{\bar 3}_F),2,0,\lambda\lambda]} &=& 5.4^{+2.3}_{-1.9} \mbox{ MeV} \, .
\end{eqnarray}
These two mass values as well as their difference are well consistent with the LHCb and CMS measurements~\cite{Aaij:2019amv,Sirunyan:2020gtz}, so our results support the interpretation of the $\Lambda_b(6146)^0$ and $\Lambda_b(6152)^0$ as $D$-wave bottom baryons of $J^P = 3/2^+$ and $5/2^+$ respectively, both of which contain two $\lambda$-mode excitations. We call it $\lambda\lambda$-mode in the present study, and its relevant multiplet is the bottom baryon doublet $[\mathbf{\bar 3}_F, 2, 0, \lambda\lambda]$.

This conclusion is the same as~\cite{Liang:2019aag,Wang:2019uaj,Chen:2019ywy,Azizi:2020tgh}, so faces the same serious problem: The lower state $\Lambda_b(3/2^+)$ would decay both into the $P$-wave $\Sigma_b \pi$ channel and the $P$-wave $\Sigma_b^* \pi$ channel, while the higher state $\Lambda_b(5/2^+)$ would dominantly decay only into the $P$-wave $\Sigma_b^* \pi$ channel, which behaviors are just opposite to the $\Lambda_b(6146)^0$ and $\Lambda_b(6152)^0$ observed by LHCb~\cite{Aaij:2019amv}, as stated in the introduction (in other words, ``face a serious problem of mass reverse''~\cite{Wang:2019uaj});

\item The masses of $\Lambda_b(5/2^+)$ and $\Lambda_b(7/2^+)$ calculated using the $[\mathbf{\bar 3}_F, 3, 1, \rho\lambda]$ multiplet are:
\begin{eqnarray}
\nonumber m_{\Lambda_b(5/2^+)} &=& 6.42^{+0.15}_{-0.11} \mbox{ GeV} \, , \,
\\ m_{\Lambda_b(7/2^+)} &=& 6.43^{+0.15}_{-0.11} \mbox{ GeV} \, , \,
\\ \nonumber \Delta m_{[\Lambda_b(\mathbf{\bar 3}_F),3,1,\rho\lambda]} &=& 14.6^{+8.8}_{-6.2} \mbox{ MeV} \, ,
\end{eqnarray}
These two mass values as well as their difference are all significantly larger than, but not too far from, the LHCb and CMS measurements~\cite{Aaij:2019amv,Sirunyan:2020gtz}. 

The advantage of this assignment is: The lower state $\Lambda_b(5/2^+)$ would dominantly decay only into the $P$-wave $\Sigma_b^* \pi$ channel, while the higher state $\Lambda_b(7/2^+)$ would decay both into the $F$-wave $\Sigma_b \pi$ channel and the $F$-wave $\Sigma_b^* \pi$ channel, which behaviors are consistent with the $\Lambda_b(6146)^0$ and $\Lambda_b(6152)^0$ observed by LHCb~\cite{Aaij:2019amv}. Note that $\sqrt{M_{\Lambda_b(6146)^0}^2 - M_{\Sigma_b}^2} \approx \sqrt{M_{\Lambda_b(6152)^0}^2 - M_{\Sigma_b}^2} \approx \sqrt{M_{\Lambda_b(6152)^0}^2 - M_{\Sigma_b^*}^2} \approx 2$~GeV, so that the $F$-wave decay widths might not be suppressed too much. Anyway, we still need to explicitly study their decay properties to verify this possibility;

\item The masses of $\Lambda_b(1/2^+)$ and $\Lambda_b(3/2^+)$ calculated using the $[\mathbf{\bar 3}_F, 1, 1, \rho\lambda]$ multiplet are:
\begin{eqnarray}
\nonumber m_{\Lambda_b(1/2^+)} &=& 6.13^{+0.10}_{-0.09} \mbox{ GeV} \, , \,
\\ m_{\Lambda_b(3/2^+)} &=& 6.13^{+0.10}_{-0.09} \mbox{ GeV} \, , \,
\\ \nonumber \Delta m_{[\Lambda_b(\mathbf{\bar 3}_F),1,1,\rho\lambda]} &=& -1.5^{+0.6}_{-0.5} \mbox{ MeV} \, .
\end{eqnarray}
This mass difference is smaller (negative) than the LHCb measurement~\cite{Aaij:2019amv}. Moreover, the decay behaviors of the $\Lambda_b(6146)^0$ and $\Lambda_b(6152)^0$ observed by LHCb~\cite{Aaij:2019amv} can not be well explained by this multiplet. Hence, our results do not favor the interpretation of the $\Lambda_b(6146)^0$ and $\Lambda_b(6152)^0$ as $D$-wave bottom baryons of $J^P = 1/2^+$ and $3/2^+$ belonging to the $[\mathbf{\bar 3}_F, 1, 1, \rho\lambda]$ multiplet.

\item The masses of $\Lambda_b(3/2^+)$ and $\Lambda_b(5/2^+)$ calculated using the $[\mathbf{\bar 3}_F, 2, 1, \rho\lambda]$ multiplet are:
\begin{eqnarray}
\nonumber m_{\Lambda_b(3/2^+)} &=& 6.17^{+0.12}_{-0.10} \mbox{ GeV} \, , \,
\\ m_{\Lambda_b(5/2^+)} &=& 6.19^{+0.12}_{-0.10} \mbox{ GeV} \, , \,
\\ \nonumber \Delta m_{[\Lambda_b(\mathbf{\bar 3}_F),2,1,\rho\lambda]} &=& 13.5^{+7.5}_{-5.6} \mbox{ MeV} \, .
\end{eqnarray}
This mass difference is a bit larger than the LHCb experiment~\cite{Aaij:2019amv}. Hence, our results do not favor the interpretation of the $\Lambda_b(6146)^0$ and $\Lambda_b(6152)^0$ as $D$-wave bottom baryons of $J^P = 3/2^+$ and $5/2^+$ belonging to the $[\mathbf{\bar 3}_F, 2, 1, \rho\lambda]$ multiplet;

\item The sum rule results extracted from the $[\mathbf{\bar 3}_F, 2, 0, \rho\rho]$ multiplet are a bit strange, because the Borel windows become larger as the threshold value $\omega_c$ decreases, which behavior has already been found in Figure~9 of Ref.~\cite{Chen:2016phw}. Hence, we do not use them to draw any conclusion.

\end{itemize}

Summarizing the above analyses, our results obtained using the method of QCD sum rules within the heavy quark effective theory support to interpret the $\Lambda_b(6146)^0$ and $\Lambda_b(6152)^0$ as $D$-wave bottom baryons of $J^P = 3/2^+$ and $5/2^+$, respectively. They both contain two $\lambda$-mode excitations, and belong to the bottom baryon doublet $[\mathbf{\bar 3}_F, 2, 0, \lambda\lambda]$. This doublet contains two other bottom baryons, $\Xi_b(3/2^+)$ and $\Xi_b(5/2^+)$, whose masses are extracted to be:
\begin{eqnarray}
\nonumber m_{\Xi_b(3/2^+)} &=& 6.26^{+0.11}_{-0.14} \mbox{ GeV} \, , \,
\\ m_{\Xi_b(5/2^+)} &=& 6.26^{+0.11}_{-0.14} \mbox{ GeV} \, , \,
\\ \nonumber \Delta m_{[\Xi_b(\mathbf{\bar 3}_F),2,0,\lambda\lambda]} &=& 4.5^{+1.9}_{-1.5} \mbox{ MeV} \, .
\end{eqnarray}
This conclusion is the same as~\cite{Liang:2019aag,Wang:2019uaj,Chen:2019ywy,Azizi:2020tgh}, but it can not well explain the decay behaviors of the $\Lambda_b(6146)^0$ and $\Lambda_b(6152)^0$ observed by LHCb~\cite{Aaij:2019amv}, as discussed above.

To solve this problem, we investigate another possible assignment, that is to interpret the $\Lambda_b(6146)^0$ and $\Lambda_b(6152)^0$ as $D$-wave bottom baryons of $J^P = 5/2^+$ and $7/2^+$ respectively, both of which belong to the $[\mathbf{\bar 3}_F, 3, 1, \rho\lambda]$ multiplet. The advantage of this assignment is that the decay behaviors of the $\Lambda_b(6146)^0$ and $\Lambda_b(6152)^0$ observed by LHCb~\cite{Aaij:2019amv} can be well explained, as discussed above. However, this assignment faces another serious problem: The masses of the $\Lambda_b(5/2^+)$ and $\Lambda_b(7/2^+)$ as well as their mass splitting are calculated in the present study to be significantly larger than, although not too far from, those of the $\Lambda_b(6146)^0$ and $\Lambda_b(6152)^0$ measured by LHCb and CMS~\cite{Aaij:2019amv,Sirunyan:2020gtz}. It is still required to explicitly study their decay properties to verify this possibility.

There exist many possible assignments for the $\Lambda_b(6072)^0$ observed by CMS and LHCb~\cite{Sirunyan:2020gtz,Aaij:2020rkw}, such as the $\Lambda_b(2S)$ state, while another possible assignment is to interpret it as the $D$-wave $\Lambda_b$ state. To verify this, one good choice is to further examine whether it has a nearby partner state in future CMS, EIC, and LHCb experiments.

To end this paper, we note that just investigating the mass spectra is not enough, and in order to well understand the $\Lambda_b(6146)^0$ and $\Lambda_b(6152)^0$~\cite{Aaij:2019amv} as well as the $\Lambda_b(6072)^0$~\cite{Sirunyan:2020gtz}, we still need to systematically study their decay properties. We have done this systematically for $P$-wave heavy baryons using the method of light-cone sum rules with the heavy quark effective theory~\cite{Chen:2017sci,Cui:2019dzj,Yang:2019cvw,Chen:2020mpy} and  are now doing this systematically for $D$-wave heavy baryons. The parameters obtained in the present study are necessary inputs.

\section*{ACKNOWLEDGMENTS}

We thank Xian-Hui Zhong for useful discussions.
This project is supported by
the National Natural Science Foundation of China under Grant No.~11722540,
the Fundamental Research Funds for the Central Universities,
and
the Foundation for Young Talents in College of Anhui Province (Grant No.~gxyq2018103).

\appendix

\section{Other sum rules}
\label{app:sumrule}

In this appendix we list the sum rules for the $[\mathbf{\bar 3}_F,2,0,\rho\rho]$, $[\mathbf{\bar 3}_F,2,0,\lambda\lambda]$, $[\mathbf{\bar 3}_F,1,1,\rho\lambda]$, $[\mathbf{\bar 3}_F,2,1,\rho\lambda]$, and $[\mathbf{\bar 3}_F,3,1,\rho\lambda]$ multiplets. We refer to~\cite{Chen:2016phw,Mao:2017wbz} for more discussions. Note that the sum rule equations for the $[\mathbf{\bar 3}_F,2,0,\rho\rho]$ and $[\mathbf{\bar 3}_F,2,0,\lambda\lambda]$ multiplets are the same as those obtained in~\cite{Chen:2016phw,Mao:2017wbz}; the equations for the $[\mathbf{\bar 3}_F,1,1,\rho\lambda]$ multiplet are different since we have explicitly used some projection operators in the present study; the equations for the $[\mathbf{\bar 3}_F,2,1,\rho\lambda]$ and $[\mathbf{\bar 3}_F,3,1,\rho\lambda]$ multiplets were not extracted in~\cite{Chen:2016phw,Mao:2017wbz}.

\begin{widetext}
The sum rule equations obtained using the interpolating field $J^{\alpha}_{3/2,+,\mathbf{\bar 3}_F,2,0,\rho\rho}$ belonging to $[\mathbf{\bar 3}_F,2,0,\rho\rho]$ are
\begin{eqnarray}
\Pi_{\Lambda_b,2,0,\rho\rho} = f_{\Lambda_b,2,0,\rho\rho}^{2} e^{-2 \bar \Lambda_{\Lambda_b,2,0,\rho\rho} / T}
&=& \int_{0}^{\omega_c} [\frac{5}{145152\pi^4}\omega^9 - \frac{5\langle g_s^2 GG \rangle}{1728 \pi^4} \omega^5]e^{-\omega/T}d\omega \, ,
\\ f_{\Lambda_b,2,0,\rho\rho}^{2} K_{\Lambda_b,2,0,\rho\rho} e^{-2 \bar \Lambda_{\Lambda_b,2,0,\rho\rho} / T}
&=& \int_{0}^{\omega_c} [ - \frac{41}{6386688\pi^4}\omega^{11} + \frac{59\langle g_s^2 GG \rangle}{90720\pi^4} \omega^7]e^{-\omega/T}d\omega \, ,
\\ f_{\Lambda_b,2,0,\rho\rho}^{2} \Sigma_{\Lambda_b,2,0,\rho\rho} e^{-2 \bar \Lambda_{\Lambda_b,2,0,\rho\rho} / T}
&=& \int_{0}^{\omega_c} [\frac{\langle g_s^2 GG \rangle}{24192\pi^4} \omega^7]e^{-\omega/T}d\omega \, ,
\end{eqnarray}
and
\begin{eqnarray}
&& \Pi_{\Xi_b,2,0,\rho\rho} = f_{\Xi_b,2,0,\rho\rho}^{2} e^{-2 \bar \Lambda_{\Xi_b,2,0,\rho\rho} / T}
\\ \nonumber &=& \int_{2m_s}^{\omega_c} [ \frac{5}{145152\pi^4} \omega^9 - \frac{m_s^2}{672\pi^4} \omega^7 - \frac{m_s\langle \bar q q \rangle}{72 \pi^2} \omega^5 + \frac{m_s\langle \bar s s \rangle}{48 \pi^2} \omega^5
\\ \nonumber && ~~~~ -\frac{5\langle g_s^2 GG \rangle}{1728 \pi^4} \omega^5 + \frac{5m_s^2\langle g_s^2 GG \rangle}{192 \pi^4} \omega^3 - \frac{5m_s\langle g_s^2 GG \rangle\langle \bar s s \rangle}{72 \pi^2} \omega ]e^{-\omega/T}d\omega \, ,
\\ && f_{\Xi_b,2,0,\rho\rho}^{2} K_{\Xi_b,2,0,\rho\rho} e^{-2 \bar \Lambda_{\Xi_b,2,0,\rho\rho} / T}
\\ \nonumber &=& \int_{2m_s}^{\omega_c} [ - \frac{41}{6386688\pi^4} \omega^{11} + \frac{197m_s^2}{483840\pi^4} \omega^9 + \frac{37m_s\langle \bar q q \rangle}{5040\pi^2} \omega^7 - \frac{277m_s\langle \bar s s \rangle}{20160\pi^2} \omega^7 - \frac{11m_s\langle g_s \bar q \sigma Gq \rangle}{180\pi^2} \omega^5
+ \frac{1921\langle g_s^2 GG \rangle}{2903040\pi^4} \omega^7
\\ \nonumber && ~~~~ - \frac{7169m_s^2\langle g_s^2 GG \rangle}{552960\pi^4} \omega^5 - \frac{13m_s\langle g_s^2 GG \rangle\langle \bar q q \rangle}{216\pi^2} \omega^3 + \frac{2381m_s\langle g_s^2 GG \rangle\langle \bar s s \rangle}{20736\pi^2} \omega^3 + \frac{121m_s\langle g_s^2 GG \rangle\langle g_s \bar q \sigma Gq \rangle}{1728\pi^2} \omega]e^{-\omega/T}d\omega\, ,
\\ && f_{\Xi_b,2,0,\rho\rho}^{2} \Sigma_{\Xi_b,2,0,\rho\rho} e^{-2 \bar \Lambda_{\Xi_b,2,0,\rho\rho} / T}
\\ \nonumber &=& \int_{2m_s}^{\omega_c} [ \frac{\langle g_s^2 GG \rangle}{24192 \pi^4} \omega^7 - \frac{m_s^2\langle g_s^2 GG \rangle}{1536 \pi^4} \omega^5 + \frac{5m_s\langle g_s^2 GG \rangle\langle \bar s s \rangle}{864 \pi^2} \omega^3]e^{-\omega/T}d\omega\, .
\end{eqnarray}

The sum rule equations obtained using the interpolating field $J^{\alpha}_{3/2,+,\mathbf{\bar 3}_F,2,0,\lambda\lambda}$ belonging to $[\mathbf{\bar 3}_F,2,0,\lambda\lambda]$ are
\begin{eqnarray}
\Pi_{\Lambda_b,2,0,\lambda\lambda} = f_{\Lambda_b,2,0,\lambda\lambda}^{2} e^{-2 \bar \Lambda_{\Lambda_b,2,0,\lambda\lambda} / T}
&=&-\int_{0}^{\omega_c} [-\frac{5}{145152\pi^4}\omega^9 + \frac{\langle g_s^2 GG \rangle}{1728\pi^4} \omega^5]e^{-\omega/T}d\omega \, ,
\\ f_{\Lambda_b,2,0,\lambda\lambda}^{2} K_{\Lambda_b,2,0,\lambda\lambda} e^{-2 \bar \Lambda_{\Lambda_b,2,0,\lambda\lambda} / T}
&=& -\int_{0}^{\omega_c} [ \frac{127}{10644480\pi^4}\omega^{11} + \frac{\langle g_s^2 GG \rangle}{17280\pi^4} \omega^7]e^{-\omega/T}d\omega \, ,
\\ f_{\Lambda_b,2,0,\lambda\lambda}^{2} \Sigma_{\Lambda_b,2,0,\lambda\lambda} e^{-2 \bar \Lambda_{\Lambda_b,2,0,\lambda\lambda} / T}
&=& -\int_{0}^{\omega_c} [-\frac{\langle g_s^2 GG \rangle}{24192\pi^4} \omega^7]e^{-\omega/T}d\omega \, ,
\end{eqnarray}
and
\begin{eqnarray}
&& \Pi_{\Xi_b,2,0,\lambda\lambda} = f_{\Xi_b,2,0,\lambda\lambda}^{2} e^{-2 \bar \Lambda_{\Xi_b,2,0,\lambda\lambda} / T}
\\ \nonumber &=& -\int_{2m_s}^{\omega_c} [\frac{m_s^2}{672\pi^4}\omega^7-\frac{5}{145152\pi^4}\omega^9 + \frac{5m_s\langle g_s^2 GG \rangle\langle \bar s s \rangle}{216\pi^2} \omega-\frac{5m_s^2\langle g_s^2 GG \rangle}{576\pi^4} \omega^3
\\ \nonumber && ~~~~~~ + \frac{\langle g_s^2 GG \rangle}{1728\pi^4} \omega^5 + \frac{m_s\langle \bar q q \rangle}{72\pi^2} \omega^5- \frac{m_s\langle \bar s s \rangle}{48\pi^2} \omega^5]e^{-\omega/T}d\omega \, ,
\\ && f_{\Xi_b,2,0,\lambda\lambda}^{2} K_{\Xi_b,2,0,\lambda\lambda} e^{-2 \bar \Lambda_{\Xi_b,2,0,\lambda\lambda} / T}
\\ \nonumber &=& -\int_{2m_s}^{\omega_c} [ - \frac{307m_s^2}{483840\pi^4}\omega^9 + \frac{127}{10644480\pi^4}\omega^{11}+ \frac{13m_s\langle g_s^2 GG \rangle\langle \bar q q \rangle}{648\pi^2} \omega^3- \frac{67m_s\langle g_s^2 GG \rangle\langle \bar s s \rangle}{2304\pi^2} \omega^3
\\ \nonumber && ~~~~~~ +\frac{1019m_s^2\langle g_s^2 GG \rangle}{184320\pi^4} \omega^5
+\frac{\langle g_s^2 GG \rangle}{17280\pi^4} \omega^7-\frac{37m_s\langle \bar q q \rangle}{5040\pi^2} \omega^7+\frac{233m_s\langle \bar s s \rangle}{20160\pi^2} \omega^7]e^{-\omega/T}d\omega \, ,
\\ && f_{\Xi_b,2,0,\lambda\lambda}^{2} \Sigma_{\Xi_b,2,0,\lambda\lambda} e^{-2 \bar \Lambda_{\Xi_b,2,0,\lambda\lambda} / T}
\\ \nonumber &=& -\int_{2m_s}^{\omega_c} [-\frac{5m_s\langle g_s^2 GG \rangle\langle \bar s s \rangle}{864\pi^2} \omega^3+\frac{m_s^2\langle g_s^2 GG \rangle}{1536\pi^4} \omega^5-\frac{\langle g_s^2 GG \rangle}{24192\pi^4} \omega^7]e^{-\omega/T}d\omega \, .
\end{eqnarray}

The sum rule equations obtained using the interpolating field $J_{1/2,+,\mathbf{\bar 3}_F,1,1,\rho\lambda}$ belonging to $[\mathbf{\bar 3}_F,1,1,\rho\lambda]$ are
\begin{eqnarray}
\Pi_{\Lambda_b,1,1,\rho\lambda} = f_{\Lambda_b,1,1,\rho\lambda}^{2} e^{-2 \bar \Lambda_{\Lambda_b,1,1,\rho\lambda} / T}
&=&-\int_{0}^{\omega_c} [-\frac{1}{16128\pi^4}\omega^9 + \frac{\langle g_s^2 GG \rangle}{512\pi^4} \omega^5]e^{-\omega/T}d\omega \, ,
\\ f_{\Lambda_b,1,1,\rho\lambda}^{2} K_{\Lambda_b,1,1,\rho\lambda} e^{-2 \bar \Lambda_{\Lambda_b,1,1,\rho\lambda} / T}
&=& -\int_{0}^{\omega_c} [ \frac{29}{1520640\pi^4}\omega^{11} - \frac{443\langle g_s^2 GG \rangle}{580608\pi^4} \omega^7]e^{-\omega/T}d\omega \, ,
\\ f_{\Lambda_b,1,1,\rho\lambda}^{2} \Sigma_{\Lambda_b,1,1,\rho\lambda} e^{-2 \bar \Lambda_{\Lambda_b,1,1,\rho\lambda} / T}
&=& -\int_{0}^{\omega_c} [\frac{\langle g_s^2 GG \rangle}{48384\pi^4} \omega^7]e^{-\omega/T}d\omega \, ,
\end{eqnarray}
and
\begin{eqnarray}
&& \Pi_{\Xi_b,1,1,\rho\lambda} = f_{\Xi_b,1,1,\rho\lambda}^{2} e^{-2 \bar \Lambda_{\Xi_b,1,1,\rho\lambda} / T}
\\ \nonumber &=& -\int_{2m_s}^{\omega_c} [-\frac{1}{16128\pi^4}\omega^9+\frac{m_s^2}{336\pi^4}\omega^7- \frac{m_s\langle \bar s s \rangle}{16\pi^2} \omega^5 +\frac{m_s\langle \bar q q \rangle}{24\pi^2} \omega^5+ \frac{\langle g_s^2 GG \rangle}{512\pi^4} \omega^5
\\ \nonumber && ~~~~~~ - \frac{65m_s^2\langle g_s^2 GG \rangle}{2304\pi^4} \omega^3- \frac{5m_s\langle g_s \bar q \sigma Gq \rangle}{24\pi^2} \omega^3+ \frac{65m_s\langle g_s^2 GG \rangle\langle \bar s s \rangle}{576\pi^2} \omega]e^{-\omega/T}d\omega \, ,
\\ && f_{\Xi_b,1,1,\rho\lambda}^{2} K_{\Xi_b,1,1,\rho\lambda} e^{-2 \bar \Lambda_{\Xi_b,1,1,\rho\lambda} / T}
\\ \nonumber &=& -\int_{2m_s}^{\omega_c} [ \frac{29}{1520640\pi^4}\omega^{11}- \frac{179m_s^2}{145152\pi^4}\omega^9 - \frac{443\langle g_s^2 GG \rangle}{580608\pi^4} \omega^7- \frac{m_s\langle \bar q q \rangle}{48\pi^2} \omega^7+\frac{167m_s\langle \bar s s \rangle}{4032\pi^2} \omega^7
+\frac{2287m_s^2\langle g_s^2 GG \rangle}{110592\pi^4} \omega^5
\\ \nonumber && ~~~~~~ +\frac{3m_s\langle g_s \bar q \sigma Gq \rangle}{16\pi^2} \omega^5+\frac{25m_s\langle g_s^2 GG \rangle\langle \bar q q \rangle}{432\pi^2} \omega^3-\frac{425m_s\langle g_s^2 GG \rangle\langle \bar s s \rangle}{2592\pi^2} \omega^3-\frac{5m_s\langle g_s^2 GG \rangle\langle g_s \bar q \sigma Gq \rangle}{288\pi^2} \omega]e^{-\omega/T}d\omega \, ,
\\ && f_{\Xi_b,1,1,\rho\lambda}^{2} \Sigma_{\Xi_b,1,1,\rho\lambda} e^{-2 \bar \Lambda_{\Xi_b,1,1,\rho\lambda} / T}
\\ \nonumber &=& -\int_{2m_s}^{\omega_c} [\frac{\langle g_s^2 GG \rangle}{48384\pi^4} \omega^7+\frac{m_s^2\langle g_s^2 GG \rangle}{2304\pi^4} \omega^5-\frac{5m_s\langle g_s^2 GG \rangle\langle \bar s s \rangle}{432\pi^2} \omega^3]e^{-\omega/T}d\omega \, .
\end{eqnarray}

The sum rule equations obtained using the interpolating field $J_{1/2,+,\mathbf{\bar 3}_F,2,1,\rho\lambda}$ belonging to $[\mathbf{\bar 3}_F,2,1,\rho\lambda]$ are
\begin{eqnarray}
\Pi_{\Lambda_b,2,1,\rho\lambda} = f_{\Lambda_b,2,1,\rho\lambda}^{2} e^{-2 \bar \Lambda_{\Lambda_b,2,1,\rho\lambda} / T}
&=&-\int_{0}^{\omega_c} [-\frac{\langle g_s^2 GG \rangle}{144\pi^4}\omega^5 + \frac{5}{36288\pi^4} \omega^9]e^{-\omega/T}d\omega \, ,
\\ f_{\Lambda_b,2,1,\rho\lambda}^{2} K_{\Lambda_b,2,1,\rho\lambda} e^{-2 \bar \Lambda_{\Lambda_b,2,1,\rho\lambda} / T}
&=& -\int_{0}^{\omega_c} [ \frac{217453\langle g_s^2 GG \rangle}{78382080\pi^4}\omega^7 - \frac{163}{3592512\pi^4} \omega^{11}]e^{-\omega/T}d\omega \, ,
\\ f_{\Lambda_b,2,1,\rho\lambda}^{2} \Sigma_{\Lambda_b,2,1,\rho\lambda} e^{-2 \bar \Lambda_{\Lambda_b,2,1,\rho\lambda} / T}
&=& -\int_{0}^{\omega_c} [\frac{17\langle g_s^2 GG \rangle}{54432\pi^4} \omega^7]e^{-\omega/T}d\omega \, ,
\end{eqnarray}
and
\begin{eqnarray}
&& \Pi_{\Xi_b,2,1,\rho\lambda} = f_{\Xi_b,2,1,\rho\lambda}^{2} e^{-2 \bar \Lambda_{\Xi_b,2,1,\rho\lambda} / T}
\\ \nonumber &=& -\int_{2m_s}^{\omega_c} [-\frac{185m_s\langle g_s^2 GG \rangle\langle \bar s s \rangle}{648\pi^2}\omega+\frac{185m_s^2\langle g_s^2 GG \rangle}{2592\pi^4}\omega^3 + \frac{25m_s\langle g_s \bar q \sigma Gq \rangle}{54\pi^2} \omega^3-\frac{\langle g_s^2 GG \rangle}{144\pi^4} \omega^5
\\ \nonumber && ~~~~~~ - \frac{5m_s\langle \bar q q \rangle}{54\pi^2} \omega^5 + \frac{5m_s\langle \bar s s \rangle}{36\pi^2} \omega^5- \frac{5m_s^2}{756\pi^4} \omega^7+\frac{5}{36288\pi^4} \omega^9]e^{-\omega/T}d\omega \, ,
\\ && f_{\Xi_b,2,1,\rho\lambda}^{2} K_{\Xi_b,2,1,\rho\lambda} e^{-2 \bar \Lambda_{\Xi_b,2,1,\rho\lambda} / T}
\\ \nonumber &=& -\int_{2m_s}^{\omega_c} [\frac{385m_s\langle g_s^2 GG \rangle\langle g_s \bar q \sigma Gq \rangle}{7776\pi^2}\omega - \frac{835m_s\langle g_s^2 GG \rangle\langle \bar q q \rangle}{5832\pi^2}\omega^3+ \frac{13693m_s\langle g_s^2 GG \rangle\langle \bar s s \rangle}{34992\pi^2} \omega^3- \frac{9277m_s^2\langle g_s^2 GG \rangle}{186624\pi^4} \omega^5
\\ \nonumber && ~~~~~~ -\frac{107m_s\langle g_s \bar q \sigma Gq \rangle}{216\pi^2} \omega^5+\frac{217453\langle g_s^2 GG \rangle}{78382080\pi^4} \omega^7+\frac{61m_s\langle \bar q q \rangle}{1134\pi^2} \omega^7-\frac{170m_s\langle \bar s s \rangle}{1701\pi^2} \omega^7
\\ \nonumber && ~~~~~~ +\frac{5749m_s^2}{1959552\pi^4} \omega^9-\frac{163}{3592512\pi^4} \omega^{11}]e^{-\omega/T}d\omega \, ,
\\ && f_{\Xi_b,2,1,\rho\lambda}^{2} \Sigma_{\Xi_b,2,1,\rho\lambda} e^{-2 \bar \Lambda_{\Xi_b,2,1,\rho\lambda} / T}
\\ \nonumber &=& -\int_{2m_s}^{\omega_c} [\frac{65m_s\langle g_s^2 GG \rangle\langle \bar s s \rangle}{1944\pi^2} \omega^3-\frac{5m_s^2\langle g_s^2 GG \rangle}{1296\pi^4} \omega^5+\frac{17\langle g_s^2 GG \rangle}{54432\pi^4} \omega^7]e^{-\omega/T}d\omega \, .
\end{eqnarray}

The sum rule equations obtained using the interpolating field $J_{1/2,+,\mathbf{\bar 3}_F,3,1,\rho\lambda}$ belonging to $[\mathbf{\bar 3}_F,3,1,\rho\lambda]$ are
\begin{eqnarray}
\Pi_{\Lambda_b,3,1,\rho\lambda} = f_{\Lambda_b,3,1,\rho\lambda}^{2} e^{-2 \bar \Lambda_{\Lambda_b,3,1,\rho\lambda} / T}
&=&\int_{0}^{\omega_c} [-\frac{21\langle g_s^2 GG \rangle}{3200\pi^4}\omega^5 + \frac{1}{12800\pi^4} \omega^9]e^{-\omega/T}d\omega \, ,
\\ f_{\Lambda_b,3,1,\rho\lambda}^{2} K_{\Lambda_b,3,1,\rho\lambda} e^{-2 \bar \Lambda_{\Lambda_b,3,1,\rho\lambda} / T}
&=& \int_{0}^{\omega_c} [ \frac{80147\langle g_s^2 GG \rangle}{38707200\pi^4}\omega^7 - \frac{223}{9856000\pi^4} \omega^{11}]e^{-\omega/T}d\omega \, ,
\\ f_{\Lambda_b,3,1,\rho\lambda}^{2} \Sigma_{\Lambda_b,3,1,\rho\lambda} e^{-2 \bar \Lambda_{\Lambda_b,3,1,\rho\lambda} / T}
&=& \int_{0}^{\omega_c} [\frac{\langle g_s^2 GG \rangle}{4800\pi^4} \omega^7]e^{-\omega/T}d\omega \, ,
\end{eqnarray}
and
\begin{eqnarray}
&& \Pi_{\Xi_b,3,1,\rho\lambda} = f_{\Xi_b,3,1,\rho\lambda}^{2} e^{-2 \bar \Lambda_{\Xi_b,3,1,\rho\lambda} / T}
\\ \nonumber &=& \int_{2m_s}^{\omega_c} [ \frac{1}{12800\pi^4} \omega^9 - \frac{3m_s^2}{800\pi^4} \omega^7 + \frac{63m_s\langle \bar s s \rangle}{800\pi^2} \omega^5
- \frac{21m_s\langle \bar q q \rangle}{400\pi^2} \omega^5 - \frac{21\langle g_s^2 GG \rangle}{3200\pi^4} \omega^5 + \frac{21m_s\langle g_s \bar q \sigma Gq \rangle}{80\pi^2} \omega^3
\\ \nonumber && ~~~~~~ + \frac{63m_s^2\langle g_s^2 GG \rangle}{1280\pi^4}\omega^3 - \frac{63m_s\langle g_s^2 GG \rangle\langle \bar s s \rangle}{320\pi^2}\omega ]e^{-\omega/T}d\omega \, ,
\\ && f_{\Xi_b,3,1,\rho\lambda}^{2} K_{\Xi_b,3,1,\rho\lambda} e^{-2 \bar \Lambda_{\Xi_b,3,1,\rho\lambda} / T}
\\ \nonumber &=& \int_{2m_s}^{\omega_c} [ -\frac{223}{9856000\pi^4} \omega^{11} + \frac{1759m_s^2}{1209600\pi^4} \omega^9
- \frac{543m_s\langle \bar s s \rangle}{11200\pi^2} \omega^7 + \frac{17m_s\langle \bar q q \rangle}{672\pi^2} \omega^7 +\frac{80147\langle g_s^2 GG \rangle}{38707200\pi^4} \omega^7
- \frac{181m_s\langle g_s \bar q \sigma Gq \rangle}{800\pi^2} \omega^5
\\ \nonumber && ~~~~~~  - \frac{1091m_s^2\langle g_s^2 GG \rangle}{38400\pi^4} \omega^5
- \frac{307m_s\langle g_s^2 GG \rangle\langle \bar q q \rangle}{4320\pi^2}\omega^3+ \frac{779m_s\langle g_s^2 GG \rangle\langle \bar s s \rangle}{3456\pi^2} \omega^3 + \frac{59m_s\langle g_s^2 GG \rangle\langle g_s \bar q \sigma Gq \rangle}{2880\pi^2}\omega ]e^{-\omega/T}d\omega \, ,
\\ && f_{\Xi_b,3,1,\rho\lambda}^{2} \Sigma_{\Xi_b,3,1,\rho\lambda} e^{-2 \bar \Lambda_{\Xi_b,3,1,\rho\lambda} / T}
\\ \nonumber &=& \int_{2m_s}^{\omega_c} [ \frac{\langle g_s^2 GG \rangle}{4800\pi^4} \omega^7 - \frac{7m_s^2\langle g_s^2 GG \rangle}{2400\pi^4} \omega^5 + \frac{7m_s\langle g_s^2 GG \rangle\langle \bar s s \rangle}{240\pi^2} \omega^3 ]e^{-\omega/T}d\omega \, .
\end{eqnarray}
\end{widetext}

\end{document}